\documentclass[a4paper,11pt]{article}
\pdfoutput=1 

\usepackage{jheppub} 

\usepackage[T1]{fontenc} 

\usepackage[utf8]{inputenc}
\usepackage{todonotes}
\usepackage{amssymb,amsmath,amsbsy}
\usepackage{braket}
\usepackage{graphicx,color}
\newcommand{\mA}{\mathcal{A}}

\newcommand{\mcI}{\mathcal{I}}
\newcommand{\mbZ}{\mathbb{Z}}

\newcommand{\mF}{\mathcal{F}}
\newcommand{\hn}{\hat{n}}
\newcommand{\mK}{\mathcal{K}}

\newcommand{\hh}{\hat{h}}

\newcommand{\mI}{\mathbb{I}}
\newcommand{\mH}{\mathcal{H}}
\newcommand{\nn}{\nonumber}
\newcommand{\lb}{\left(}
\newcommand{\rb}{\right)}
\newcommand{\ee}{\end{equation}}
\newcommand{\bea}{\begin{eqnarray}}
\newcommand{\eea}{\end{eqnarray}}

\usepackage{slashed}
\usepackage{pifont}%
\usepackage{natbib}
\usepackage{graphicx}
\usepackage[vcentermath]{youngtab}
\usepackage{ytableau}
\usepackage{subcaption}

\title{\boldmath Generalized Entanglement Entropy, Charges, and Intertwiners}

\author[a]{Keiichiro Furuya,}
\author[a,b,1]{Nima Lashkari,\note{Corresponding author.}}
\author[a]{Shoy Ouseph}

\affiliation[a]{Department of Physics and Astronomy, Purdue University, West Lafayette, IN
47907, USA}
\affiliation[b]{School of Natural Sciences, Institute for Advanced Study, Princeton, New Jersey 08540, USA}

\emailAdd{kfuruya@purdue.edu}
\emailAdd{nima@purdue.edu}
\emailAdd{souseph@purdue.edu}

\abstract{The entanglement theory in quantum systems with internal symmetries is rich due to the spontaneous creation of entangled pairs of charge/anti-charge particles at the entangling surface. We call these pair creation operators the bi-local intertwiners because of the role they play in the representation theory of the symmetry group. We define a generalized measure of entanglement entropy as a measure of information erased under restriction to a subspace of observables. We argue that the correct entanglement measure in the presence of charges is the sum of two terms; one measuring the entanglement of charge-neutral operators, and the other measuring the contribution of the bi-local intertwiners. Our expression is unambiguously defined in lattice models as well in quantum field theory (QFT). We use the Tomita-Takesaki modular theory to highlight the differences between QFT and lattice models, and discuss an extension of the algebra of QFT that leads to a factorization of the charged modes.}

\begin{document} 
\maketitle

\flushbottom

\section{Introduction}\label{sec:1}

The study of entanglement in many-body quantum systems has opened new windows to understanding strongly coupled phenomena. Entanglement measures in lattice models have helped identify phases of matter and universal dynamical processes. In Poincare-invariant quantum field theory (QFT), entanglement measures have taught us about universal long-range correlation patterns, and renormalization monotones \cite{casini2004finite,casini2012renormalization,casini2017modular}. In holographic QFT, entanglement measures play an important role in the emergence of geometry out  of quantum states \cite{nishioka2009holographic}. In this work, we study the entanglement theory in quantum systems with conserved charges.

In the conventional quantum information theory, the Hilbert space of a bipartite system $A_{12}\equiv A_1\cup A_2$ with $A_1$ and $A_2$ non-overlapping is the tensor product of the Hilbert spaces of each: $\mH_{12}=\mH_1\otimes\mH_2$. There are local algebras of operators on $A_1$ and $A_2$ that we denote by $\mF_1$ and $\mF_2$, respectively. For instance, the algebra of operators of a $d$-level quantum system (qudit) is the algebra of $d\times d$ complex matrices. The global algebra of the bipartite system $A_{12}$ is $\mF_{12}=\mF_1\otimes \mF_2$. 
The local algebra $\mF_1$ is a subalgebra of $\mF_{12}$, and the reduced state on this subalgebra is given by a partial trace on $\mF_2$. The entanglement measure we are interested in captures the amount of information erased by partial trace. Entanglement is a resource that can be distilled in the form of Einstein-Podolosky-Rosen (EPR) pairs and can be used to teleport quantum states. For instance, for a bipartite qudit density matrix $\rho_1\otimes \rho_2$ the amount of information erased by partial trace on $A_2$ is $\log d-S_{vN}(\rho_2)$, where $S_{vN}(\rho)=-\text{tr}(\rho\log\rho)$ is the von Neumann entropy. The state $\rho_1\otimes \mI_2/d$ is unique in that it loses no information under partial trace.
The distinguishability of an arbitrary state $\rho_{12}$ with respect to the invariant state of partial trace $\rho_1\otimes \mI_2/d$ can be used to quantify the amount of information lost in partial trace of $A_2$. In quantum information theory, the distinguishability of a state $\rho$ from $\sigma$ is measured by the relative entropy 
\begin{eqnarray}\label{relative}
S(\rho\|\sigma)=\text{tr}(\rho\log\rho)-\text{tr}(\rho\log\sigma)
\end{eqnarray}
which is non-negative and vanishes if and only if $\rho=\sigma$.
We choose the relative entropy $S(\rho_{12}\|\rho_1\otimes \mI_2/d)=\log d-S_{vN}(\rho_{12})+S_{vN}(\rho_1)$ as our measure of the information lost in partial trace.\footnote{It has an operational interpretation in the language of the state merging protocol \cite{horodecki2007quantum}.}

In systems with symmetries and conserved charges, the degrees of freedom in $A_1$ and $A_2$ are not completely independent. Charge conservation requires that any physical process that creates a charge particle in $A_1$ also creates the opposite charge in $A_2$. If we superpose states of different charge, there is no information in their relative phase because they cannot be detected in any physical process made out of charge conserving operations. 
The naive relative entropy for a charged system cannot be used as a  resource to distill entangled pairs \cite{gour2009measuring,bartlett2007reference}. In this work, we argue that the measure of entanglement with the correct operational interpretation is the sum of two relative entropies. One term captures the entanglement due to the charge-neutral degrees of freedom. These operators are invariant under the symmetry transformation. The second term captures the contribution of charged operators, and is a measure of the asymmetry of states in the resource theory of symmetry \cite{gour2009measuring,bartlett2007reference}. In section \ref{sec:2}, we motivate a generalized entanglement entropy beyond the case of tensor products, and connect it to the coarse-grained entropy defined by the Jaynes maximum entropy principle \cite{jaynes1957information}. For other definitions of generalized entanglement see \cite{barnum2004subsystem,barnum2003generalizations}.

\begin{figure}[t]
 \centering
 \begin{subfigure}{.4\textwidth}
   \centering
   \includegraphics[width=.5\linewidth]{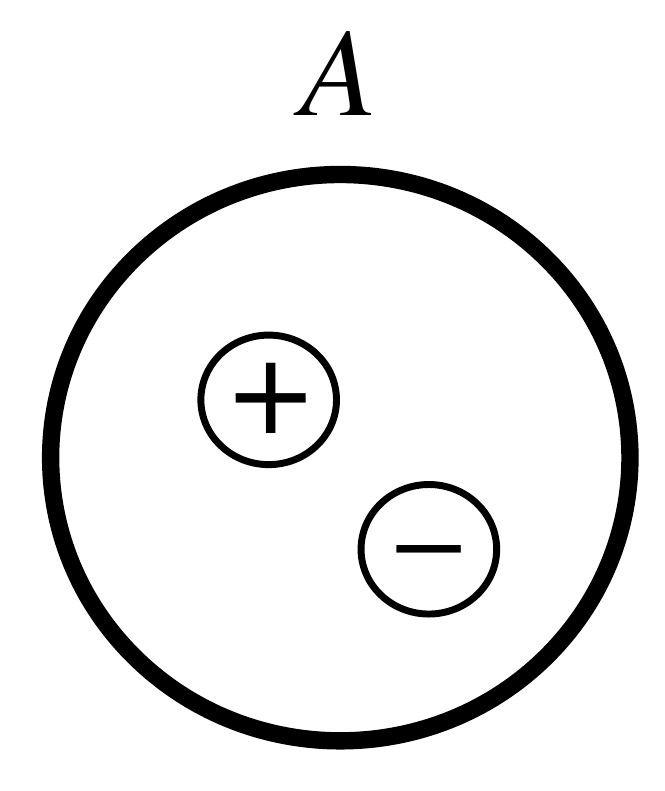}
   \caption{}
   \label{fig1:sub1}
 \end{subfigure}
 \begin{subfigure}{.4\textwidth}
   \centering
   \includegraphics[width=.5\linewidth]{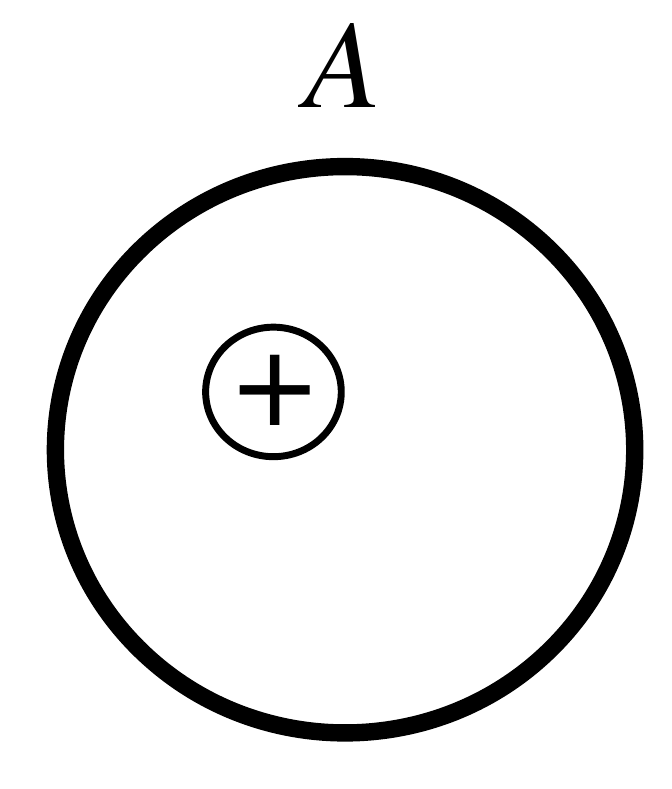}
   \caption{}
   \label{fig1:sub2}
 \end{subfigure}
 \caption{\small{ a) A charge neutral operator in region $A$: $a\in \mA_A$. b) A charged operator in region $A$: $b\in\mF_A$.}}
 \label{fig1}
\end{figure}

\begin{figure}[t]
 \centering
 \begin{subfigure}{.3\textwidth}
   \centering
   \includegraphics[width=.8\linewidth]{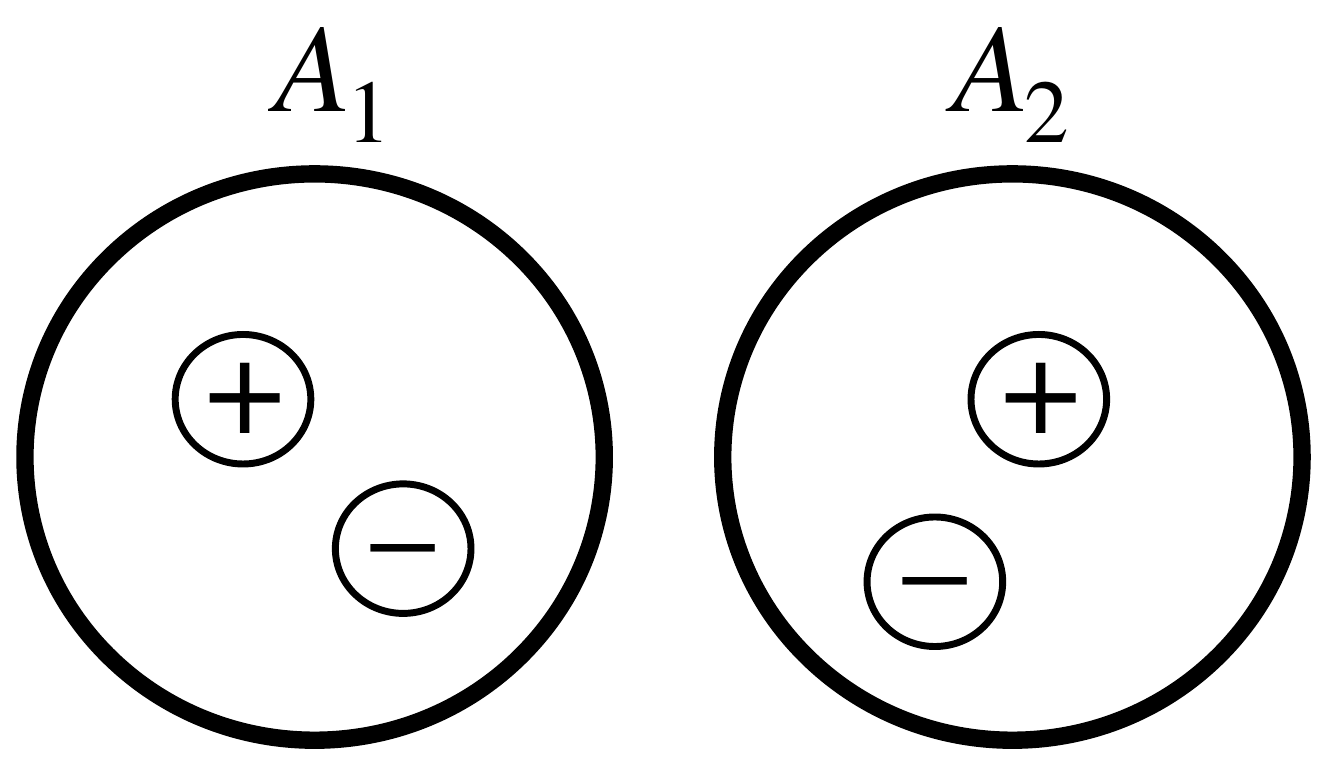}
   \caption{}
   \label{fig2:sub1}
 \end{subfigure}
 \begin{subfigure}{.3\textwidth}
   \centering
   \includegraphics[width=.8\linewidth]{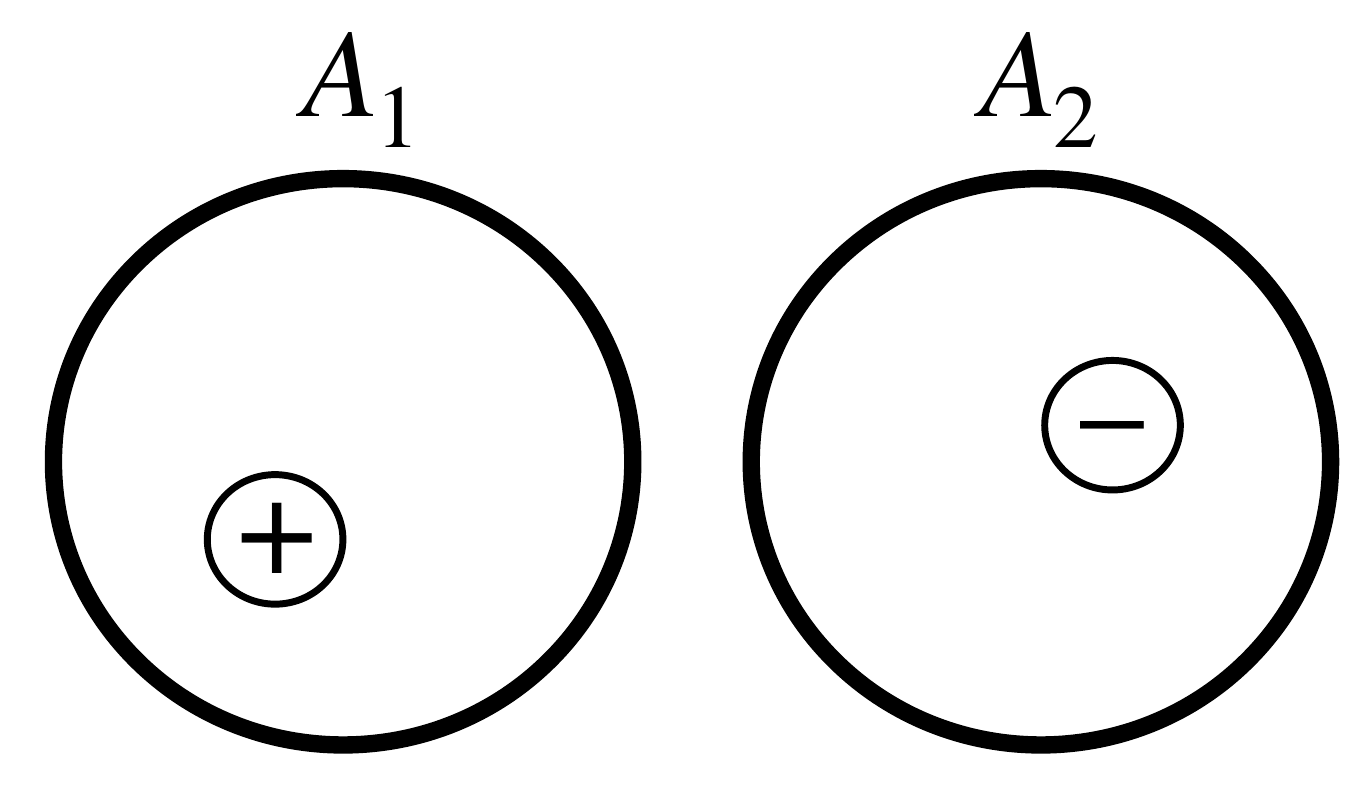}
   \caption{}
   \label{fig2:sub2}
 \end{subfigure}
 \begin{subfigure}{.3\textwidth}
   \centering
   \includegraphics[width=.8\linewidth]{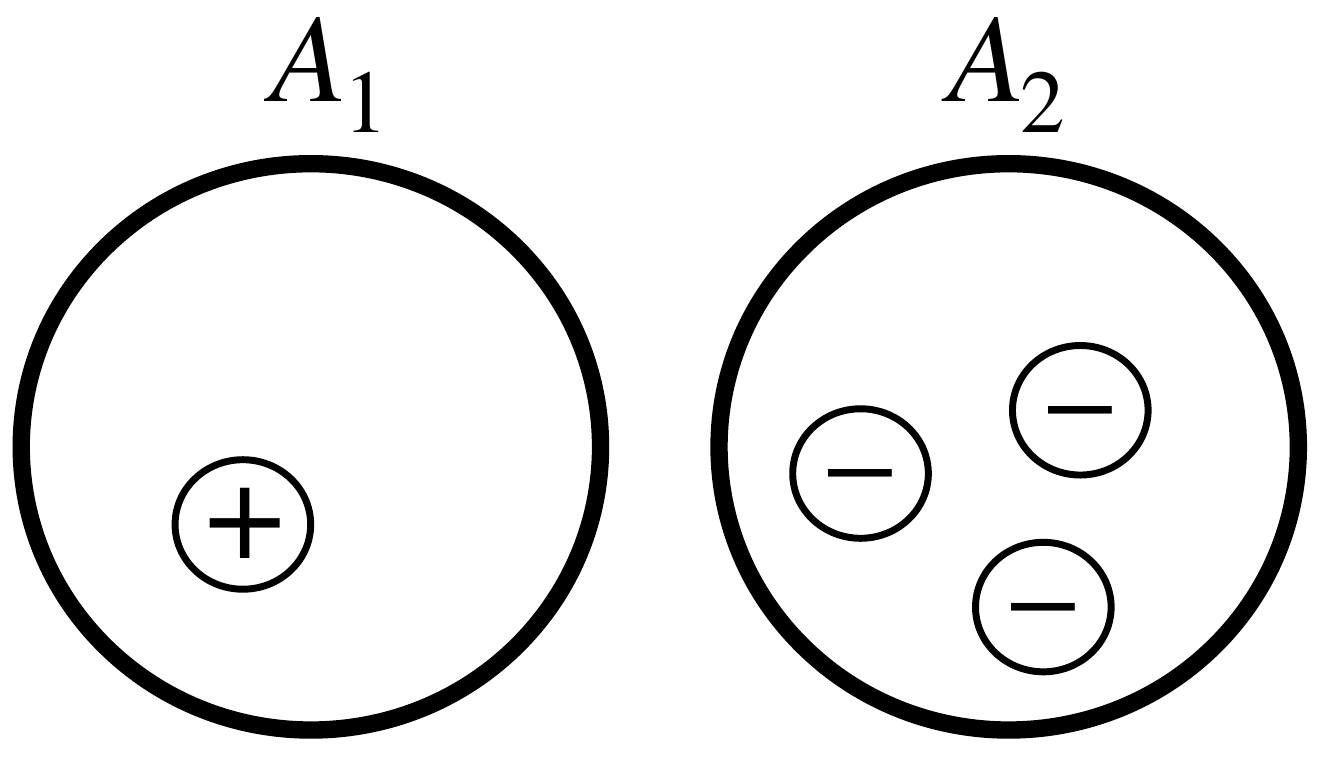}
   \caption{}
   \label{fig2:sub3}
 \end{subfigure}
 \caption{\small{(a) Charge neutral operators in region $A_1 \cup A_2$: $a\in \mA_1 \otimes \mA_2$. (b) A bi-local intertwiner in $A_1 \cup A_2$: $\mathcal{I}_{12} \in \mA_{12}$. (c) Local intertwiners, or charged operators in $A_1 \cup A_2$ that belongs to the global algebra $\mF_{1} \otimes \mF_{2}$.}}
 \label{fig2}
\end{figure}

The charge-neutral operators in $\mF$ form a sub-algebra that we denote by $\mA$; figure \ref{fig1}. In the bipartite setup, the algebra of charge-neutral operators localized in $A_1$ is a subalgebra of all charge-neutral operators of $A_{12}$: $\mA_1\subset \mA_{12}$. 
However, it is not true that $\mA_1$ and $\mA_2$ generate all the charge-neutral operators of $\mA_{12}$. The operators that spontaneously create a pair of charge particle in $A_1$ and its anti-charge in $A_2$ belong to $\mA_{12}$, but not to $\mA_1\otimes \mA_2$. In section \ref{sec:3}, we call such operators {\it bi-local intertwiners} due to the role they play in the representation theory of the symmetry group; see figure \ref{fig2}. Our goal is to quantify the contribution of the local intertwiners to the entanglement.
The key idea is to associate to any state $\rho$ an invariant state $E^*(\rho)$. The expectation value of all charge-neutral operators $\mA$ in $E^*(\rho)$ and $\rho$ match, however the probability for the spontaneous creation of a charge/anti-charge pair in the invariant state is zero. The relative entropy $S(\rho\|E^*(\rho))$ measures the distinguishability of the two states. It is a measure of the asymmetry of $\rho$ and captures the information contained in the bi-local intertwiners. In section \ref{sec:2}, we argue that this relative entropy added to the mutual information between region $A_1$ and $A_2$ due to the charge-neutral algebra $\mA_1\otimes\mA_2$ captures the total amount of entanglement between $A_1$ and $A_2$.
This quantity is also discussed in previous work of \cite{hollands2018entanglement,Casini:2019kex} and some of the ideas here parallel those of \cite{Casini:2019kex}.

In section \ref{sec:3}, we review the representation theory of symmetry groups and the superselection sectors. A special role is played by the charge creation/annihilation operators that take charge neutral operators from a superselection sector to another. They are called intertwiners and together with the charge neutral sub-algebra they generate the algebra of all charged particles. In section \ref{sec:3} we provide simple physical examples from qubits to QFT to demonstrate the formalism. A reader who is already familiar with the formalism can skip this section. In section \ref{sec:4}, we make the distinction between global algebras and local algebras. In the global case, we consider the algebra of charge neutral operators as a sub-algebra of all charged operators $\mA\subset \mF$. In the local case, we consider the tensor product of charge neutral operators in non-overlapping regions $A_1$ and $A_2$ as a sub-algebra of charge-neutral operators of $A_1\cup A_2$: $\mA_1\otimes \mA_2\subset \mA_{12}$.

The study of entanglement in QFT is subtle due to absence of a tensor product $\mH_A\otimes \mH_{A'}$ that reflects itself as ultra-violet divergence in the entanglement entropy \cite{calabrese2004entanglement,hollands2018entanglement,witten2018aps}. Modular theory is a mathematical framework that is well-suited for the study of entanglement in any quantum system from qubits to QFT. In modular theory, instead of tensor products and local density matrices, the algebra of operators localized in a region and locality constraints among them are used to define entanglement measures. In section \ref{sec:5}, we use modular theory to define both the relative entropies that measure the entanglement between non-touching regions $A_1$ and $A_2$ in a QFT with conserved charges. We highlight the difference in the analysis of entanglement between QFTs and lattice models. Finally, we discuss an extension of the QFT algebra that factors out charged excitations and brings the QFT algebra closer to lattice models.

In this work, we focus on global symmetries, however, the formalism can be generalized to many gauge theories \cite{buchholz1982locality,haag2012local}. We postpone this to future work.

\section{Generalizations of entanglement}\label{sec:2}

\subsection{Conditional expectation as generalization of partial trace}\label{sec:2.1}

Consider the algebra of operators of two qudits $\mF_{12}=\mF_1\otimes \mF_2$ and the subalgebra of operators localized on the first system $\mF_1\otimes \mathbb{I}_2$. 
The reduced density matrix on $\mF_1$ is given by the partial trace over $\mF_2$: $\rho_1=\text{tr}_2(\rho_{12})$.
In the classical case, $\rho_{12}=\sum_{kk'}p_{kk'}\ket{kk'}\bra{kk'}$ the reduced density matrix on the first qudit is $\rho_1=\sum_kq_k\ket{k}\bra{k}$ where $q_k=\sum_{k'} p_{kk'}$ are the classical conditional expectations to obtain result $k$ in a measurement on first qudit: $q_k=\text{tr}\lb (\ket{k}\bra{k}\otimes \mathbb{I})\rho_{12}\rb$. In a mathematical analogy, one can think of density matrices as non-commutative probabilities and partial trace as non-commutative conditional expectation \cite{petz2007quantum}. 

\begin{figure}[h]
 \centering
 \begin{subfigure}{1\textwidth}
   \centering
   \includegraphics[width=.7\linewidth]{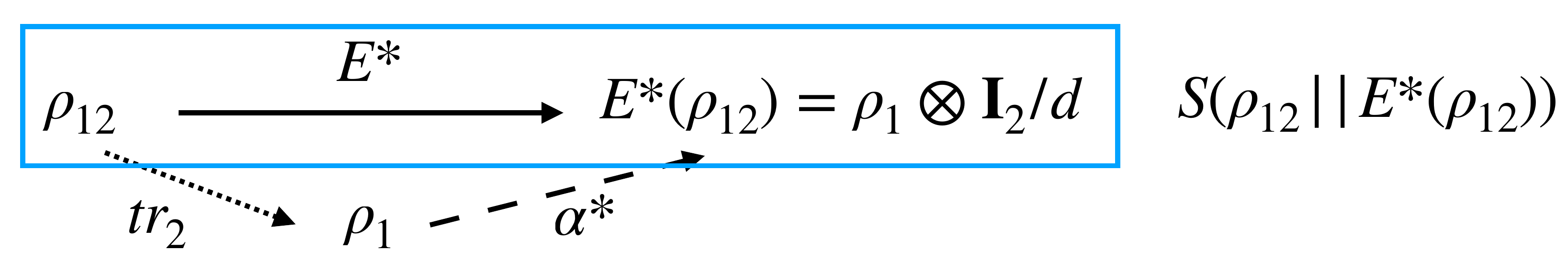}
   \caption{}
   \label{fig3:sub1}
 \end{subfigure}
 \begin{subfigure}{1\textwidth}
   \centering
   \includegraphics[width=.7\linewidth]{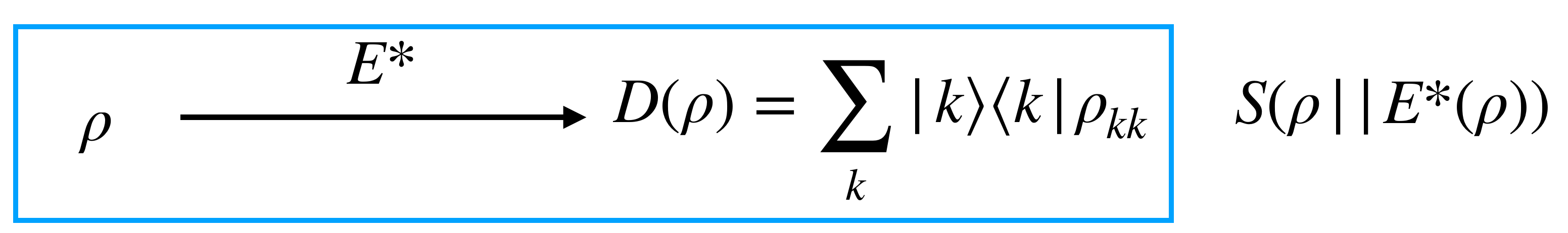}
   \caption{}
   \label{fig3:sub2}
 \end{subfigure}
 \caption{\small{Our entanglement measure is the relative entropy of the state $\rho$ with respect to its corresponding invariant state $E^*(\rho)$: $S(\rho\|E^*(\rho))$. (a) The example where the map $E^*$ is a composition of partial trace of system $2$ and the recovery map $\alpha^*$ which results in an invariant state $E^*(\rho_{12})$. (b) The example where the map $E^*$ decoheres the density matrix $\rho$ in a particular basis $\{\ket{k}\}$.}}
 \label{fig3}
\end{figure}

To compute how much information was erased during partial trace we have to pull $\rho_1$ back to the bipartite Hilbert space by a linear map that we denote by $\alpha^*(\rho_1)=\phi_{12}$ with the following properties:
\begin{enumerate}
    \item It is consistent with $\rho_1$: $\text{tr}_2(\phi_{12})=\rho_1$.

    \item The state $\phi_{12}$ is invariant under partial trace and $\alpha^*$: $\alpha^*(\phi_1)=\phi_{12}$ so that $\alpha^*$ does not add any information.
\end{enumerate} 
We call such $\alpha^*$ maps {\it recovery maps} or state extensions \cite{petz2007quantum,lashkari2019entanglement}. In the partial trace case, the recovery map with the properties above is $\alpha^*(\rho_1)=\rho_1\otimes \mI_2/d$.\footnote{An example of a map that satisfies the first property but not the second is $\alpha^*(\rho_1)=\rho_1\otimes \omega_2$ for some $\omega_2$.} 
It is convenient to think of partial trace and recovery together as one linear map that sends density matrices on $\mF_{12}$ to the density matrices on the subalgebra $\mF_1\otimes \mathbb{I}_2$: $E^*(\rho_{12})=\rho_1\otimes \mI_2/d$. 
The dual of the $E^*$ is a projection from $\mF_{12}$ down to the subalgebra $\mF_1\otimes \mI_2$:
\begin{eqnarray}
E(b_1\otimes b_2)=b_1\otimes \mI_2\:\text{tr}\lb \frac{b_2}{d}\rb\ .
\end{eqnarray}
Here, by the duality we mean going from the Schrodinger to the Heisenberg picture\footnote{An alternative notation used in \cite{Casini:2019kex} is to denote $E^*(\rho)$ by $\rho\circ E$.}  
\begin{eqnarray}
 \text{tr}(E^*(\rho_{12}) (b_1\otimes b_2))=\text{tr}(\rho_{12} E(b_1\otimes b_2))=\text{tr}(\rho_1 b_1)\:\text{tr}\lb \frac{b_2}{d}\rb\ .
\end{eqnarray}
The map $E$ has the property that it squares to itself, i.e. $E^2=E$, so that $E^*(\rho_{12})$ is invariant state of $E$:
\begin{eqnarray}
\text{tr}(E^*(\rho_{12})E(b_1\otimes b_2))=\text{tr}\lb E^*(\rho_{12}) (b_1\otimes b_2)\rb\ . 
\end{eqnarray}
The relative entropy of $\rho_{12}$ with respect to the invariant state $E^*(\rho_{12})=\rho_1\otimes \mI_2/d$ measures the asymmetry of the state or the amount of information erased in partial trace: $S(\rho_{12}\|E^*(\rho_{12}))\geq 0$; see figure \ref{fig3}.

 A simple way to generalize partial trace is to consider a more general dual map $E:\mF_{12}\to \mF_1\otimes\mI_2$:
 \begin{eqnarray}\label{mD}
 &&E(b_1\otimes b_2)=b_1\otimes \mathbb{D}(b_2)\nn\\
 &&\mathbb{D}(b)=\sum_k b_{kk}\ket{k}\bra{k}
\end{eqnarray}
where $\{\ket{k}\}$ is some distinguished basis of the second qudit.
 In the Schrodinger picture, the state transforms according to
\begin{eqnarray}\label{invariantCond}
&&E^*(\rho_{12})=\sum_k p_k \rho_1^{(k)}\otimes \ket{k}_2\bra{k}_2\nn\\
&&p_k\rho_1^{(k)}=\bra{k}_2\rho_{12}\ket{k}_2,\qquad p_k=\text{tr}(\rho_{12}(\mathbb{I}\otimes \ket{k}_2\bra{k}_2))
\end{eqnarray}
that dephases the density matrix and erases the information in the off-diagonal operators $\ket{k}\bra{k'}$.
Similar to the $E$ of partial trace we have the property that $E(\mI)=\mI$ so that $E^*(\rho)$ is properly normalized. Furthermore, $E$ squares to itself which implies that $E^*(\rho_{12})$ is an invariant state of $E$.

In systems with conserved charges, the subalgebra of charge-neutral operators corresponds to matrices that are block-diagonal in some basis labelled by charge.
For instance, take a qubit and the symmetry transformation $\sigma_z$. The Abelian subalgebra $\mathcal{D}\subset \mF$ of $2\times 2$ complex matrices diagonal in $\sigma_z$ basis is the charge neutral algebra. 
The dephasing map $E(b)=\mathbb{D}(b)$ projects operators from $\mF$ to $\mathcal{D}$. 
For a general quantum system with symmetry we need to define a linear map $E:\mF\to\mA$ with $\mA\subset \mF$ the subalgebra of charge-neutral operators as a generalization of partial trace. An example of one such maps is the Haar average over the group $G$:
\begin{eqnarray}\label{HaarAverage}
E(b)=\frac{1}{|G|}\int dg U_g^\dagger b U_g\ .
\end{eqnarray}
The operator $E(b)$ is charge-neutral for any charged operator $b$.
In analogy to partial trace, we require this map preserves the identity operator, and it leaves the charge-neutral operators unchanged so that the state $E^*(\rho)$ defined by $(E^*(\rho))(b)=\rho(E(b))$ for all $b\in\mF$ is invariant under the map $E$: $E(E^*(\rho))=E^*(\rho)$. 
The generalization of partial trace is called the {\it non-commutative  conditional expectation} (or in short conditional expectation) that is a linear map from $\mF$ to an arbitrary subalgebra $\mA$ such that $E(\mI)=\mI$ and $E(ab)=a E(b)$ for all $a\in\mA$ and $b\in\mF$  \cite{petz2007quantum,ohya2004quantum}.\footnote{In this paper, the operator $b$ is chosen to belong to the algebra of charged operators, whereas $a$ denotes a charge-neutral operator.}
Since $E(a)=a$ all invariant operators are in $\mA$ and every operator in $\mA$ is invariant. As a result, $E^*(\rho)\in\mA$.

\subsection{Generalized entanglement entropy and coarse-grained entropy}\label{sec:2.2}

In conventional quantum information theory, the amount of entanglement between $A_1$ and $A_2$ is measured by the distinguishability of the $\rho_{12}$ with respect to the unentangled state $\rho_1\otimes \rho_2$:
\begin{equation}
    S(\rho_{12}\|\rho_1\otimes \rho_2)=-S_{vN}(\rho_{12})+S_{vN}(\rho_1)+S_{vN}(\rho_2)
\end{equation}
which is called the mutual information.
Consider a multi-partite global state $\ket{\Omega}_{AA'}$ and its reduced states  $\rho_A$ and $\rho_{A'}$ on region $A$ and the complementary region $A'$, respectively. The distinguishability of $\ket{\Omega}$ from the tensor product state $\rho_A\otimes\rho_{A'}$ is measured by the relative entropy 
\begin{eqnarray}
S(\ket{\Omega}\bra{\Omega}\|\rho_A\otimes\rho_{A'})=2S_{vN}(\rho_A)\ .
\end{eqnarray} 
The tensor product state $\rho_A\otimes \rho_{A'}$ has the same expectation values as $\ket{\Omega}$ for all operators in $\mF_A\otimes \mathbb{I}$ and $\mathbb{I}\otimes\mF_{A'}$, however, all correlations between $A$ and $A'$ are erased. The expectation of all operators $b\otimes b'$ with $b\in \mF_A$ and $b'\in\mF_{A'}$ factors in the tensor product state $\rho_A\otimes\rho_{A'}$.

To generalize the notion of entanglement to a general subalgebra $\mA\subset \mF$ we invoke the Jaynes maximum entropy principle. Consider the set of all density matrices $\sigma$ that have the same expectation values as $\rho$ for operators in $\mA$: $\text{tr}((\sigma-\rho) a)=0$ for all $a\in\mA$. According to Jaynes the entropy of a state $\rho$ with respect to a subalgebra $\mA$ is the supremum of the von Neumann entropy $S_{vN}(\sigma)$ over the set of all consistent states $\sigma$ \cite{jaynes1957information}:
\begin{eqnarray}\label{Jaynessubalgebra}
S_J(\rho,\mA)=S_{vN}(\sigma_{max})
\end{eqnarray}
where $\sigma_{\max}$ is consistent with $\rho$ and has the maximum entropy. 
Hereafter, we suppress the $vN$ index of the von Neumann entropy.

The Jaynes maximum entropy consistent state is precisely our invariant state $E^*(\rho)$.
Given a general conditional expectation $E$ and a state $\sigma$ consistent with $\rho$ on $\mA$ we have 
\begin{eqnarray}
\text{tr}((E^*(\sigma)-E^*(\rho))b)=\text{tr}((\sigma-\rho)E(b))=0,
\end{eqnarray}
therefore $E^*(\sigma)=E^*(\rho)$. At the end of section \ref{sec:2.1} we showed that the invariant state is in $\mA$, therefore the logarithm of an invariant state is also in $\mA$:
\begin{eqnarray}
\text{tr}(\sigma \log E^*(\rho))&=&\text{tr}(\sigma  E(\log E^*(\rho)))=\text{tr}(E^*(\sigma)\log E^*(\rho))\nn\\
&=&\text{tr}(E^*(\rho)\log E^*(\rho))=-S(E^*(\rho))\ .
\end{eqnarray}
In the above, we have assumed that the conditional expectation preserves the trace: $\text{tr}(E(b)-b)=0$ \cite{Casini:2019kex}.\footnote{We thank Horacio Casini for pointing this out to us.}
From the definition (\ref{relative}) it follows that the relative entropy of any consistent state $\sigma$ consistent with $\rho$ on $A$ with respect to the invariant state $E^*(\rho)$ is
\begin{eqnarray}\label{relentropyinv}
S(\sigma\|E^*(\rho))&=&-S(\sigma)+S(E^*(\rho))\geq 0\ .
\end{eqnarray}
From the positivity of relative entropy we conclude that the invariant state of a conditional expectation $E$ is the maximum entropy state appearing in the Jaynes formula:
\begin{eqnarray}
E^*(\rho)=\sigma_{max}
\end{eqnarray}
and the non-degeneracy of relative entropy tells us that this state is unique.\footnote{If $\sigma_{max}$ and $\sigma'_{max}$ are both maximum entropy then $S(\sigma_{max}\|\sigma'_{max})=0$, therefore $\sigma_{max}=\sigma'_{max}$.}
Therefore, our proposed measure of the information lost in $E$ is the entanglement deficit from the maximum value:
\begin{eqnarray}
S(\rho\|E^*(\rho))=S(\sigma_{max})-S(\rho)\ .
\end{eqnarray}

As an example, consider the subalgebra of matrices $\mA=\mF_1\otimes \mathcal{D}_2$ and the set of all $\sigma$ that are consistent with $\rho$ on $\mA$ and maximize the entropy among them. The consistent states are all $\sigma_{12}$ that satisfy $\text{tr}((\sigma_{12}-\rho_{12})(a_1\otimes \ket{k}\bra{k}))=0$ for all basis vectors $\ket{k}$. The relative entropy of $\sigma_{12}$ with respect to the invariant state in (\ref{invariantCond}) is
\begin{eqnarray}
S(\sigma_{12}\|\sum_k p_k \rho_1^{(k)}\otimes \ket{k}\bra{k})=-S(\sigma_{12})+H(p)+\sum_k p_k S(\rho^{(k)}_1)\geq 0
\end{eqnarray}
where $H(p)=-\sum_k p_k\log(p_k)$ is the Shannon entropy of $p_k$ \cite{nielsen2002quantum}.
The maximum entropy state is the invariant state, and the Jaynes entropy is 
\begin{eqnarray}
S_J(\rho_{12},\mF_1\otimes \mathcal{D}_2)=S(E^*(\rho_{12}))=H(p)+\sum_k p_kS(\rho_1^{(k)})\ .
\end{eqnarray}
The reduced state on system $A_1$ is $\rho_1=\sum_k p_k\rho^{(k)}_1$. The von Neumann entropy of $\rho_1$ is less than the Jaynes entropy because of the inequality \cite{nielsen2002quantum}
\begin{eqnarray}
 S(\sum_k p_k \rho_1^{(k)})\leq H(p)+\sum_kp_k S(\rho_1^{(k)})\ .
\end{eqnarray}

The definition of Jaynes entropy can be generalized beyond subalgebras to any subspace of observables $P$:
\begin{eqnarray}
S_J(\rho,P)=\sup_{\sigma\in \mF^*}\{S_{vN}(\sigma)|\text{tr}((\sigma-\rho)a)=0, \forall a\in P\}
\end{eqnarray}
where $\mF^*$ denotes the set of all states of the global algebra $\mF$. This measure is often called the {\it coarse-grained entropy}.
For instance, consider the subspace of observables built out of linear sums of $a_1\otimes \mathbb{I}$ and $\mathbb{I}\otimes a_2$ and a bipartite density matrix $\rho_{12}$.
The relative entropy $S(\sigma_{12}\|\rho_1\otimes \rho_2)=S(\rho_1)+S(\rho_2)-S(\sigma_{12})=I(1:2)\geq 0$, where $I(1:2)$ is the mutual information between site one and two. Therefore, the maximum entropy state in the Jaynes formula that reduces to both $\rho_1$ and $\rho_2$ is $\rho_1\otimes \rho_2$ and as a result
$S_J(\rho_{12},P)=S(\rho_1\otimes \rho_2)=S(\rho_1)+S(\rho_2)$ \cite{lashkari2019entanglement}. 
Our relative entropy measure
\begin{eqnarray}
&&S(\rho_{12}\|\sigma_{max})=S(\rho_{12}\|\rho_1\otimes\rho_2)=I(1:2)
\end{eqnarray}
equals the mutual information that well
captures the amount of correlations between $A_1$ and $A_2$.
In the absence of a subalgebra and a conditional expectation $\sigma_{max}$ replaces $E^*(\rho)$ and we propose $S(\rho\|\sigma_{max})$ as a measure of the information lost under restriction to the subspace of observables $P$ \cite{lashkari2019entanglement}.
To find the maximum entropy state consider the Lagrange multipliers $\lambda_i$ and the function 
\begin{eqnarray}
-\text{tr}(\sigma\log\sigma)+\sum_i \lambda_i\text{tr}((\rho-\sigma)\mathcal{O}_i)
\end{eqnarray}
where $\mathcal{O}_i$ is a basis for the subspace of observables $P$.  Setting the variation of the expression above with respect to $\sigma$ and $\lambda_i$ establishes that the maximum entropy state $\log\sigma_{max}=\sum_i \mu_i \mathcal{O}_i\in P$ for some constants $\mu_i$.
Similar to the case of conditional expectation the maximum entropy state belongs to the subspace $P$, i.e. $\sigma_{max}\in P$, and the expectation value of every operator that is not in $P$ is zero. As a result
\begin{eqnarray}
S(\rho\|\sigma_{max})=-S(\rho)-\text{tr}(\rho\log\sigma_{max})=S(\sigma_{max})-S(\rho)\ .
\end{eqnarray}

In QFT the von Neumann entropy of a region is divergent\footnote{It is a property of the algebra and not the states.} and we can only compute the relative entropy of states. This motivates us to replace Jayne's maximum entropy principle with the supremum of $S(\rho\|\sigma)$ over all $\sigma$ consistent with $\rho$ on $P$:
\begin{eqnarray}\label{GenEntropy}
 I_P(\rho)=\sup_{\sigma\in \mF^*}\{ S(\rho\|\sigma)| \text{tr}((\sigma-\rho)a)=0, \forall a\in P\}
\end{eqnarray}
that is the measure of information entropy produced under the restriction to a subspace of observables $P$ and has the advantage of being well-defined in QFT like in systems with density matrices.
We postpone further discussion of the generalized entanglement to future work and in the remainder of this work focus on the case of charge-neutral subalgebras.

In a system with an internal symmetry group $G$, the symmetry transformation acts on the local algebra of region $A$ as a unitary transformation: $b_i\to U_g^\dagger b_i U_g$ for all $b_i\in \mF_i$ and $U_g$ some unitary representation of $G$.
The operators in $\mF_i$ that are invariant under the action of the symmetry form a subalgebra of uncharged operators that we denote by $\mA_i$: 
\begin{eqnarray}
U_g^\dagger a_i U_g=a_i,\forall a_i\in\mA_i\ .
\end{eqnarray}
On a lattice, there is a unitary operator localized in $\mF_i$ that acts the same way as $U_g$  on $\mF_i$:
\begin{eqnarray}
\tau_g b_i \tau_g^\dagger=U_g b_i U_g^\dagger,\qquad \forall b_i\in \mF_i
\end{eqnarray}
we call this operator the {\it twist} and it generates another representation of the group that we call the {\it twist group} $G_\tau$: $\tau_g \tau_h=\tau_{gh}$.
The commutator of the twist with the group action is
\begin{eqnarray}
U_g \tau_h U_g^\dagger=\tau_{gh g^{-1}}\ .
\end{eqnarray}
For instance, in a bipartite system with symmetry transformation $U_g=e^{i g (Q_1+Q_2)}$ where $Q_1+Q_2$ is the total charge of $A_{12}$ the twist is $\tau_g=e^{i g Q_1}$; see figure \ref{fig2}. It belongs to $\mF_1$ and acts the same way as $U_g$ on $\mF_1$. We postpone the subtleties in defining $\tau_g$ in QFT to section \ref{sec:5}. The algebra $\mA_{12}$ of charge-neutral operators in $A_{12}$ is larger than the algebra generated by locally charge-neutral operators of $A_1$ and $A_2$, namely $\mA_1\otimes \mA_2$. This is because there are operators that correspond to the creation of a pair of charged particles of opposite charge one in region $A_1$ and the other in $A_2$. We call these operators the bi-local intertwiners $\mathcal{I}_{12}$. 
We will see in section \ref{sec:4} that there exists a conditional expectation constructed from the twist group $E_\tau:\mA_{12}\to \mA_1\otimes \mA_2$ that washes out the information content of the bi-local intertwiners: $E_\tau(\mathcal{I}_{12})=0$.\footnote{The map $E_\tau$ is from $\mF_{12}$ to $\mA_1\otimes \mA_2$. However, we will be mostly concerned with its action on $\mA_{12}$.} The amplitude for the invariant state $E_\tau^*(\rho_{12})$ to spontaneously create an entangled pair of charge/anti-charge particles is zero. The relative entropy $S(\rho_{12}\|E_\tau^*(\rho
_{12}))$ measures the amount of correlations due to the bi-local intertwiners. Note that the reduced state on $\mA_1\otimes \mA_2$ still contains lots of correlations in between region one and two. It is only the correlations due to intertwiners that are washed out. 
In the presence of charges, the naive mutual information  $S^{\mF_{12}}(\rho_{12}\|\rho_1\otimes \rho_2)$ contains unphysical correlations that cannot be accessed in any charge-conserving process. We would like to discard all operators that create charge on $A_{12}$. First, we restrict the relative entropy to the invariant algebra $\mA_{12}$. In general, the relative entropy $S^{\mA}(\rho\|\omega)$ is a measure of distinguishability of the two states using only the operators in $\mA$. Alternatively, one can think of this relative entropy as
\begin{eqnarray}
S^{\mA_{12}}(\rho\|\omega)=S^{\mF_{12}}(E^*(\rho)\|E^*(\omega))
\end{eqnarray}
where $E:\mF_{12}\to \mA_{12}$. The expression above implies that the distinguishability of invariant states of $E$ does not change under the restriction to the invariant subalgebra $\mA_{12}$ \cite{petz2007quantum}.
Second, we replace $\rho_1$ with $E_\tau^*(\rho_1)$ to make sure that $E_\tau^*(\rho_1)\otimes \rho_2$ has no bi-local intertwiners.
Therefore, we consider the measure $S^{\mA_{12}}(\rho_{12}\|E_\tau^*(\rho_1\otimes \rho_2))$.

\begin{figure}[h]
 \centering
   \includegraphics[width=.5\linewidth]{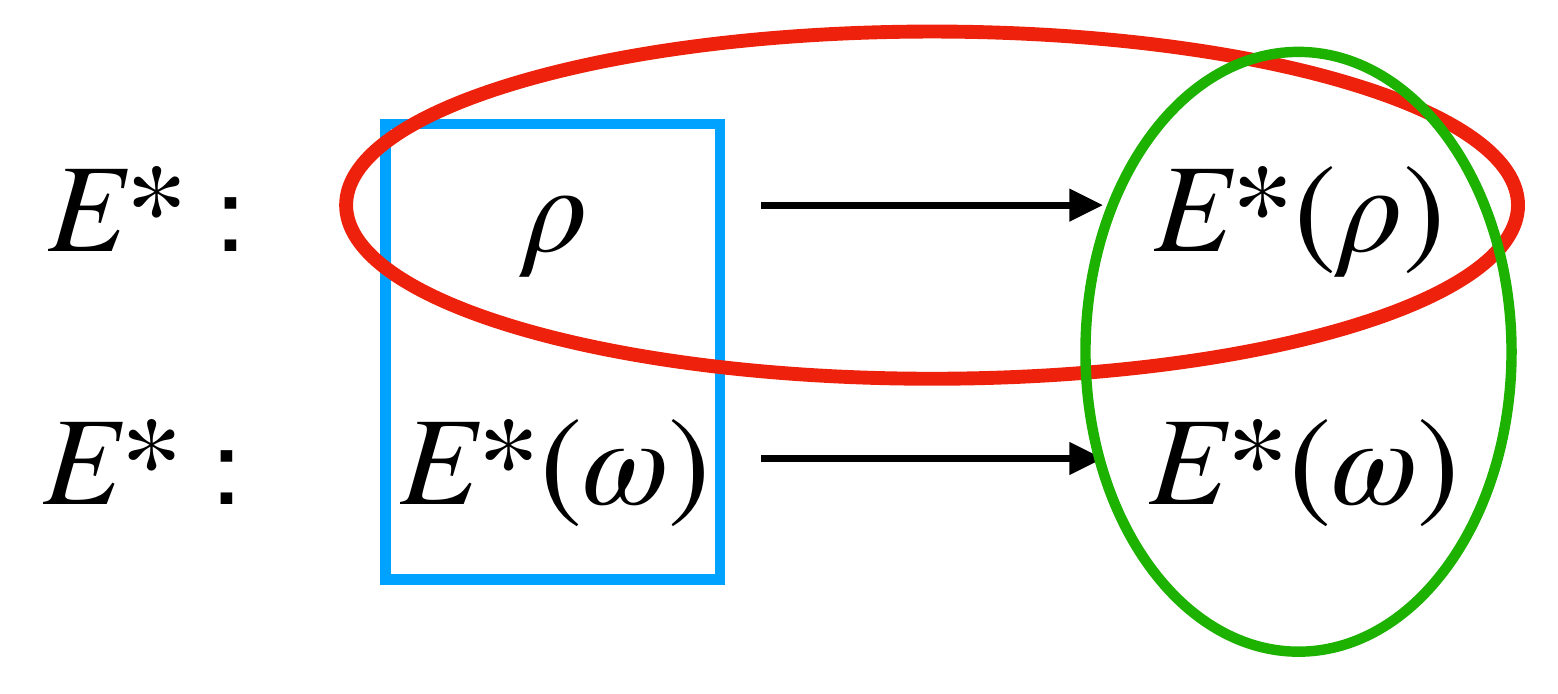}
 \caption{\small{A pictorial description of the relative entropy property in equation (\ref{thm9}) written in terms of states of $\mF$. The relative entropy of blue states $S(\rho||E^*(\omega))$ is the relative entropy the red ellipse $S(\rho\|E^*(\rho))$ plus the relative entropy of the green ellipse $S(E^*(\rho)\|E^*(\omega))$. Note that since both $E^*(\rho)$ and $E^*(\omega)$ are invariant under $E$, the green relative entropy is the same as $S^{\mA}(\rho\|E^*(\omega))$.}}
 \label{fig4}
\end{figure}

A useful property of relative entropy is that it satisfies the following equality (Theorem 9.3 of \cite{petz2007quantum}); see figure \ref{fig4}:
\begin{eqnarray}\label{thm9}
S^{\mF}(\rho\|E^*(\omega))=S^{\mA}(\rho\|E^*(\omega))+S^{\mF}(\rho\|E^*(\rho))
\end{eqnarray}
where $E:\mF\to \mA$. Applying the identity above to the twist conditional expectation $E_\tau$ implies that our measure splits into two terms \footnote{This identity was also used in \cite{longo2018relative} to compute relative entropies in QFT.}
\begin{eqnarray}\label{final1}
S^{\mA_{12}}(\rho_{12}\|E_\tau^*(\rho_1)\otimes \rho_2))=S^{\mA_1\otimes \mA_2}(\rho_{12}\|E_\tau^*(\rho_1)\otimes \rho_2)+S^{\mA_{12}}(\rho_{12}\|E_\tau^*(\rho_{12}))\ .
\end{eqnarray}
The first term is the relative entropy with respect to the charge-neutral operators of $A_1$ and $A_2$, and the second term is the contribution due to the bi-local intertwiners.
We can use the conditional expectation $E:\mF_{12}\to \mA_{12}$ to rewrite both terms in terms of the charged algebras:
\begin{eqnarray}\label{final2}
&&S^{\mF_{12}}(E^*(\rho_{12})\|E^*(E_\tau^*(\rho_{12})))+S^{\mF_{12}}(E^*(E_\tau^*(\rho_{12}))\|E^*(E_\tau^*(\rho_1)\otimes \rho_{2}))\ .
\end{eqnarray}
In section \ref{sec:3}, we will see that the conditional expectations $E$ and $E_\tau$ are Haar averages over the group and the twist group, respectively. If $\rho_{12}$ is invariant under $U_g$ we get the following simplification 
\begin{eqnarray}\label{entanglementinv}
S^{\mF_{12}}(\rho_{12}\|E_\tau^*(\rho_{12}))+S^{\mF_{12}}(E_\tau^*(\rho_{12})\|\rho_1\otimes \rho_{2})\ .
\end{eqnarray}

From the conditional expectation in (\ref{HaarAverage}) it is clear that our relative entropies have the general form $S(\sum_k p_k\rho_k\|\sum_k q_k\omega_k)$. Relative entropy satisfies the inequality
\begin{eqnarray}\label{convexity}
S(\sum_k p_k\rho_k\|\sum_k q_k\omega_k)\leq H(p\|q)+\sum_k p_k S(\rho_k\|\omega_k)
\end{eqnarray}
where $H(p\|q)$ is the classical Kullback–Leibler divergence of the probability distributions $p_k$ and $q_k$.
To see this, consider the block-diagonal density matrices $\rho=\oplus_k p_k\rho_k$ and $\omega=\oplus_k q_k \omega_k$ the relative entropy 
\begin{eqnarray}\label{relentropyadd}
S(\rho\|\omega)=H(p\|q)+\sum_k p_k S(\rho_k\|\omega_k)\ .
\end{eqnarray}
With respect to the subalgebra of operators $\mI\otimes a$ the density matrix is $\sum_k p_k \rho_k$ and $\sum_k q_k\omega_k$ and since relative entropy is monotonic under restriction to the subalgebra we find that relative entropy satisfies (\ref{convexity}). In section \ref{sec:5}, we generalize this inequality to QFT and use it to bound the relative entropies in (\ref{entanglementinv}) from above and below.

\section{Symmetry and intertwiners}\label{sec:3}

\subsection{Superselection sectors and intertwiners}

We start by reviewing some definitions and set the notations for our discussion of quantum systems with symmetries.
Consider a quantum system and its Hilbert space $\mH$. The set of all bounded linear operators acting on this Hilbert space forms an algebra, $B(\mH)$, that acts irreducibly on $\mH$. We call this algebra the field algebra and denote it by $\mathcal{F}$. All proper subalgebras of $\mF$ act reducibly on $\mH$. A symmetry is a linear transformation of operators in the algebra
$b\to \alpha_g(b)\in \mF$ that respects operator multiplication: $\alpha_g(b_1b_2)=\alpha_g(b_1)\alpha_g(b_2)$ and is invertible.\footnote{In mathematical language, such a transformation is called an automorphism of the algebra. If we relax the invertibility assumption we have an endomorphism of the algebra.}
The set of all symmetry transformations of the algebra forms the symmetry group $G$. By Wigner's theorem, any symmetry is represented by either a unitary or anti-unitary transformation of the Hilbert space, i.e. $\ket{\Psi}\to U_g\ket{\Psi}$, and acts on the algebra as $\alpha_g(b)=U_g^\dagger b U_g$. The set of operators $a$ that commute with $U_g$ form a subalgebra $\mA\subset \mF$ that we refer to either as the invariant subalgebra, or the subalgebra of charge-neutral operators. On a lattice if the group $G$ is Abelian $U_g$ is itself charge neutral and belongs to $\mA$.\footnote{When $U_g$ is not in $\mA$ we say the symmetry transformation is an outer automorphism of the algebra $\mA$.}

If there exist vectors in the Hilbert space such that 
\begin{eqnarray}
\braket{\Phi|U_g|\Psi}=0
\end{eqnarray}
for all $U_g\in G$ we say that $\ket{\Phi}$ and $\ket{\Psi}$ belong to different {\it selection} sectors.\footnote{If there exists no selection sectors; that is to say the only subspace of $\mH$ invariant under the symmetry transformation is the whole $\mH$ we say the action of the symmetry is {\it ergodic}. For instance, the action of modular flow on local algebras of QFT is ergodic.} The Hilbert space splits into a direct sum of selection sectors $\mH=\oplus_r \mK_r\otimes \mH_r$ where $\mK_r$ is the irreducible representation $r$ of $G$ and $\mH_r$ is the Hilbert space corresponding to the charge neutral degrees of freedom. 
The basis of the Hilbert space is $\ket{r,i}\otimes\ket{\alpha}$  where $i=1,\cdots, d_r$ with $d_r$ the dimension of the irreducible representation $r$. The group acts as $U=\oplus_r U^{(r)}_g\otimes 1_r$, and by Schur's lemma the invariant operators of each irreducible representation are $\mathbb{I}_r\otimes a$, where $\mathbb{I}_r=\sum_{i=1}^{d_r}\ket{r,i}\bra{r,i}$ is the identity operator in the Hilbert space $\mK_r$ of representation $r$.
The subalgebra of invariant operators is $\oplus_r \mI_r\otimes a$ which has the non-trivial center $\oplus_r \lambda_r \mI_r\otimes 1_r$.
If the group $G$ is Abelian all its irreducible representations are one-dimensional and we can label them by charge $q$: $\mH=\oplus_q \ket{q}\bra{q}\otimes \mH_q$ or simply $\mH=\oplus_q \mH_q$.

Consider the Abelian group $\mbZ_d$ and its irreducible representations labelled by charge $q$: $U^q_g=e^{2\pi i g q/d}$ with $g=0,\cdots d-1$ and $q=0,\cdots d-1$. The regular representation of $G$ is the vector space $\mK$ of a qudit:
\begin{eqnarray}
U_g=\sum_{h} \ket{(g+h)\bmod d}\bra{h}
\end{eqnarray}
where $g+h$ is the group multiplication and the identity element is zero charge.
The irreducible representations are all one-dimensional and correspond to basis where all $U_g$ are diagonal
\begin{eqnarray}
&&U_g=\sum_q e^{2\pi i g q/d}\ket{q}\bra{q}\nn\\
&&\ket{q}=\sum_g e^{-2\pi i g q/d}\ket{g}\ .
\end{eqnarray}
 The {\it dual group} $\hat{G}$ is the Fourier space generated by 
 \begin{eqnarray}\label{Uq}
 \hat{U}_q=\sum_g e^{-2\pi i g q/d}\ket{g}\bra{g}=\sum_k \ket{(q+ k)\bmod d}\bra{k}\ .
 \end{eqnarray}
The elements of the dual group take us in between irreducible representations and commute with the action of the invariant subalgebra
\begin{eqnarray}\label{interAbelian}
\hat{U}_q \ket{k}\bra{k}=\ket{k+q}\bra{k+q} \hat{U}_q\ .
\end{eqnarray}
The operators that satisfy the equation above are called the intertwiners, and physically they are charge creation/annihilation operators. Take the infinite Abelian group $G=U(1)$ of rotations around a circle. The irreducible representations are constant momentum modes and the intertwiners are the operators that add momentum $\hat{U}_q=\sum_q\ket{q+k}\bra{k}$ and generate the dual group $\hat{G}=\mbZ$ with the multiplication operation that adds charges $k+q$.

Consider a finite non-Abelian group $G$ represented in its regular representation by a qudit of dimension $|G|$:
\begin{eqnarray}\label{nonAbel}
U_g=\sum_h \ket{g h}\bra{h}
\end{eqnarray}
where $gh$ is the group multiplication. The Hilbert space splits into $\mK=\oplus_{r,i}  \mK_{r,i}$ where the irreducible representation $r$ with the index $i$ running from zero to the dimension $d_r$. The irreducible representation $r$ appears $d_r$ times in the decomposition of the regular representation, therefore $\sum_r d_r^2=|G|$. 
An operator in $\mK_r$ can be written as $\sum_{ij}b_{ij}\ket{r,i}\bra{r,j}$ but by Schur's lemma the invariant operators are proportional to $\mI_r$.
The intertwiners are linear maps that take us in between different irreducible representations and commute with the action of the invariant operators in the algebra:
\begin{equation}
    V_{r,i}\mI_r=\ket{0}\bra{0}V_{r,i}\ .
\end{equation}
The partial isometry $V_{r,i}=\frac{1}{\sqrt{d_r}}\ket{0}\bra{r,i}$ satisfies this equation, and is the non-Abelian analog of $\ket{0}\bra{q}$.
The map $\rho_r$ maps operators from the charged sectors to the vacuum sector:
\begin{eqnarray}
\rho_r(\mI_r a)=\sum_i V_{r,i} \mI_r a V_{r,i}^\dagger
\end{eqnarray}
where $a\in\mathbb{C}$ is a complex number here.
In the Abelian case, we constructed a unitary $\hat{U}_k$ by adding $\ket{q+k}\bra{q}$ that generates the dual group $\hat{G}$. For an arbitrary charge-neutral operator  $\tilde{a}=\sum_q a_q\ket{q}\bra{q}$ we have
\begin{eqnarray}
&&\rho_k(\tilde{a})=\hat{U}_k^\dagger \tilde{a} \hat{U}_k\nn\\
&&\hat{U}_k=\sum_q \ket{q+k}\bra{q}
\end{eqnarray}
which is a generalization of (\ref{Uq}) to an arbitrary Abelian group.
However, in the non-Abelian case, adding a charge $r$ to another charge $r'$ corresponds to the tensor multiplication of two irreducible representations that is not irreducible. The dual $\hat{G}$ to a non-Abelian group $G$ is not a group. The elements of the dual to a non-Abelian group are different representations (not necessarily irreducible), and their multiplication is tensor multiplication but there is no inverse operation. As we will see in the next section, when the representation is infinite dimensional the operators $V_{r,i}$ can be thought of as isometries that take us between the irreducible representations.
 
If the symmetry group $G$ is compact there is a normalizable Haar measure $dg$ and we can integrate over the group to project to the zero charge sector $P_0=\ket{0}\bra{0}\otimes 1$:
\begin{eqnarray}
\frac{1}{|G|}\int_{g\in G} dg\: U_g\ket{\Psi}=P_0\ket{\Psi}
\end{eqnarray}
where $|G|$ is the volume of the group.
The resulting subspace is called the vacuum sector which is spanned by all the invariant states of $G$. 
For an Abelian group $G$ the other irreducible representations are found using a Fourier transform with $q\in \hat{G}$ with the group multiplication being the addition of charges:
\begin{eqnarray}\label{fourierirrep}
\frac{1}{|G|}\int_{g\in G} dg\:e^{\frac{2\pi i g q}{|G|}} U_g\ket{\Psi}=P_q\ket{\Psi}\ .
\end{eqnarray}
The non-Abelian analog of this projector is
\begin{eqnarray} \label{Pr}
P_r=\frac{d_r}{|G|}\int_{g\in G} dg \: \chi_r^*(g)\: U_g
\end{eqnarray}
where $\chi_r(g)$ is the character of the irreducible representation $r$.

We say two vectors $\ket{\Psi}$ and $\ket{\Phi}$ belong to different {\it superselection sectors} of algebra $\mA$ if $\braket{\Psi|a\Phi}=0$ for all $a\in\mA$. For instance, states $\ket{\Psi_q}$ and $\ket{\Phi_{q'}}$ that were in different selection sectors of $\mF$, belong to different superselection sectors of the neutral subalgebra $\mA$.
Given an algebra $\mF$ and a compact symmetry group $G$ 
the linear map $E:\mF\to\mA$ that computes the group average of an operator $b\in \mF$ is a conditional expectation to the charge-neutral subalgebra
\begin{eqnarray}
E(b)=\frac{1}{|G|}\int_{g\in G} dg \: U_g^\dagger b U_g\ .
\end{eqnarray}
because it satisfies  $E(a b)=a E(b)$ for all $a\in \mA$ and $b\in \mF$. This is the conditional expectation that we advocated in section \ref{sec:2}. 

We can reconstruct the field algebra $\mF$ from the charge-neutral subalgebra algebra $\mA_q$ by adding the intertwiners back. 
In the Abelian case, the intertwiners $\hat{U}_q=\sum_q \ket{q\pm 1}\bra{q}$ are unitaries of the dual group. They create or annihilate charges.
Enlarging the algebra of charge-neutral operators by added to it $\hat{U}_q$ and taking the closure generates the full algebra of charged operators.
In the non-Abelian case, $\hat{G}$ the dual group is mathematically not a group. However, we can still enlarge the charge-neutral algebra by adding the intertwiners to obtain the full algebra $\mF$. 
In representation theory language, enlarging the algebra $\mA$ by including intertwiners corresponds to the crossed product of $\mA$ by the dual group $\hat{G}$: $\mA\rtimes \hat{G}$, see appendix \ref{app:A} for the definition of the dual group and crossed product.

In the remainder of this section, we provide several examples of quantum systems with symmetry and highlight the role of the intertwiners. The first four examples have an Abelian symmetry group and the last two have a non-Abelian symmetry.
We postpone the discussion of intertwiners for local algebras until the next section.

\subsection{Example 1: Qudit}

Consider the Hilbert space of a qubit $\mH_2$ and the algebra of $2\times 2$ complex matrices. Take the symmetry transformation to be the group $\mbZ_2$ generated by the transformations: $\alpha_1(a)=a$ and $\alpha_g(a)=\sigma_z a\sigma_z$.\footnote{We use the notation $\mbZ_n=\mathbb{Z}/n\mathbb{Z}$.}
Here, $U_g=\sigma_z=(-1)^Q$ where $Q=\frac{1}{2}(1-\sigma_z)$ is the charge operator.
The algebra of charge neutral operators $\mathcal{D}_2$ is the algebra of matrices diagonal in the $\sigma_z$ basis. The Hilbert space splits into two sectors $\mH_0\oplus\mH_1$ with $P_q=\ket{q}\bra{q}$ projecting to the sector of charge $q$. 
The intertwiner $V=\ket{0}\bra{1}$ solves the equation (\ref{interAbelian}) and relates the two charged sectors. The dual group is the $\mbZ_2$ that is generated by $\sigma_x=V+V^\dagger$. If we add the intertwiner (or the generator of the dual group $\sigma_x$) to the invariant algebra $\mathcal{D}_2$ we obtain the full algebra of the qubit.

For a qudit the Hilbert space is spanned by $\ket{k}$ with $k=1,\cdots d$, and we take the symmetry group to be $\mbZ_d$ generated by the diagonal matrices $\sum_k e^{2\pi ig k/d}\ket{k}\bra{k}$. The invariant sub-algebras are one-dimensional $\mA_k=a\ket{k}\bra{k}$ and the projections to the superselection sectors are $P_k=\ket{k}\bra{k}$. Each $\ket{k'}\bra{k}$ is a unitary intertwiner from $\mH_k$ to $\mH_{k'}$. The dual group is the Fourier transform $\mbZ_d$ generated by the unitary $\sum_k \ket{(k+1)\mod d}\bra{k}$. 

The generalization to infinite dimension is immediate. 
Take the Hilbert space of a free particle on a circle and the rotation group around the circle: $G=U(1)$. The Hilbert space splits into one dimensional irreducible representations of the rotation group $\mH=\oplus_{k\in \mathbb{Z}}\ket{k}\bra{k}$  where $\ket{k}$ is a momentum eigenstate. The invariant algebras are $\mA_k=a\ket{k}\bra{k}$, and the intertwiners are $\ket{k'}\bra{k}$. The dual group is $\mathbb{Z}$ generated by the momentum addition/subtraction operator $\sum_k\ket{k\pm 1}\bra{k}$. 
Adding the intertwiners to the invariant algebra gives all operators in the Hilbert space of free quantum particle on a circle.

\subsection{Example 2: Non-relativistic quantum fields}

Consider a non-relativistic bosonic or fermionic field on a circle and assume that the total number of particles is conserved. The particle number operator is $N=\int dx\: a^\dagger(x)a(x)$ and the symmetry transformations are $e^{i\alpha N}$.
The Fock space is a direct sum of sectors with fixed particle number $n$: $\mH=\oplus_{n\in \mathbb{N}}\mH_n$ with vectors in each $\mH_n$ represented by totally symmetric (anti-symmetric) wave-functions of $n$-variable: $\psi_{\pm}^{(n)}(x_1,\cdots x_n)$.  The intertwiners that take us in between sectors are the creation/annihilation operators $a_{\pm}^\dagger(f)/a_\pm(f)$ that map $\mH_n$ to $\mH_{n+1}$ and back according to
\begin{eqnarray}
&&(a_{\pm}^\dagger(f)\psi^{(n)})(x_1,\cdots, x_{n+1})= \frac{1}{\sqrt{n+1}}\sum_{k=1}^{n+1}(\pm 1)^{k-1}f(x_k)\psi^{(n)}(x_1,\cdots, x_{k-1},x_{k+1},\cdots, x_{n+1})\nn\\
&&(a_{\pm}(f)\psi^{(n+1)})(x_1,\cdots ,x_n)=\sqrt{n+1}\int dy \overline{f(y)}\psi^{(n+1)}(y,x_1,\cdots, x_{n})
\end{eqnarray}
and $f$ is a bounded complex function on the circle \cite{bratteli1996operator}.
There are many intertwiners corresponding to different functions $f$, however adding one of them to the invariant algebra suffices to generate the full algebra. We choose $\int dx |f(x)|^2=1$ so that the intertwiner is an isometry: $(a(f)a^\dagger(f)\psi^{(n)})=(a^\dagger(f)a(f)\psi^{(n)})=\psi^{(n)}$.
The full algbera $\mF$ is generated by operators $a_{\pm}(f)$ and $a_{\pm}^\dagger(f)$ satisfying
\begin{eqnarray}
&&[a_\pm(f),a_\pm(g)]_{\pm}=0,\qquad [a_\pm(f),a_\pm^\dagger(g)]_\pm=\braket{f,g}\mathbb{I}\nn\\
&&[a,b]_-\equiv ab+ba,\qquad [a,b]_+=ab-ba,\qquad \braket{f,g}=\int dx \overline{f(x)}g(x)\ .
\end{eqnarray}
The dual group is generated by the field operator $\Phi(f)=a(f)+a^\dagger(f)$.

\subsection{Example 3: Free relativistic fermions}\label{sec:3.4}

In a general relativistic theory particle number is not conserved. However, in the case of free fermions the transformation $(-1)^Q$ with $Q=\int j^0(x)$ remains a symmetry, where $j^0(x)=:\Psi^\dagger(x)\Psi(x):$ is the charge density operator. The full algebra $\mF$ is generated by $\Psi(f)=\int d^2 x f(x)\:\Psi(x)$ where $f$ is a function of spacetime that solves the classical equations of motion \cite{hollands2018entanglement}. 
The Hilbert space splits into two sectors $\mH=\mH_+\oplus\mH_-$ that correpond to the even and odd number of fermions. The invariant algebra $\mA$ is generated by all the operators with an even number of fermions, e.g. $X=\Psi(y)\Psi(z)$ or $Y=\Psi(y)\Psi(z)^\dagger$.\footnote{The commutators are $[Q,X]=-2X$ and $[Q,Y]=0$.}
The operator $\Psi(f)$ adds a unit of charge and intertwines the two sectors. The unitary $\hat{U}(f)=\Psi(f)+\Psi^\dagger(f)$ with $\int dx \:|f(x)|^2=1$ generates the $\mbZ_2$ dual group: $(1,\hat{U}(f))$. It has the following properties:
\begin{eqnarray}
&&\hat{U}(f)X =X\hat{U}(f) -f(z)\Psi(y)+f(y)\Psi(z)\nn\\
&&\hat{U}(f)Y =Y\hat{U}(f)-f(z)\Psi(y)+f(y)\Psi^\dagger(z)\ . 
\end{eqnarray}
Each choice of $f$ leads to a particular choice of $\mbZ_2$. If we add any $\hat{U}(f)$ to the algebra of invariant operators all other charged operators $\hat{U}(g)$  are created by closing the algebra, because $\hat{U}(f)^\dagger \hat{U}(g)$ is charge-neutral. Representations with different values of $f$ are unitarily equivalent by the inner automorphism $\hat{U}(f)^\dagger \hat{U}(g)$.

The maps $\rho_f(a)\equiv \hat{U}(f) a\hat{U}(f)^\dagger$ are outer automorphisms of the invariant algebra $a\in\mA$:
\begin{eqnarray}
&&\rho_f(a)\in\mA \nn\\
&&\rho_f(a_1 a_2)=\rho_f(a_1)\rho_f(a_2)\ .
\end{eqnarray}
For instance, for the total charge we have
\begin{eqnarray}
\rho_f(Q)=Q+\Psi(f)\Psi^\dagger(f)-\Psi^\dagger(f)\Psi(f)\ .
\end{eqnarray}
The operator $\hat{U}(f)$ has charge one:
\begin{eqnarray}
\rho_f((-1)^Q)=-(-1)^Q
\end{eqnarray}
which implies that an average over the dual group kills the symmetry transformation
\begin{eqnarray}
(-1)^Q+\rho_f((-1)^Q)=0\ .
\end{eqnarray}

\subsection{Example 4: $U(1)$ current algebra}\label{sec:3.5}

As the next example, consider the algebra of a free compact relativistic boson in two dimensions on a circle. 
The shift of the scalar field $\phi\to \phi+a$ is a $U(1)$ global symmetry. In the radial quantization frame, we consider the algebra of $W(u)=e^{i J(u)}$ with $J(u)=\int \frac{dz}{2\pi i}J(z) u(z)$ with $u(z)$ a smooth function on the circle.
It is generated by the $U(1)$-invariant current $J(z)=(\partial\phi)(z)=\sum_{n\in \mathbb{Z}}z^{-n-1}j_n$. The scalar field expanded in terms of $j_n$ modes is
\begin{eqnarray}
\phi(z,\bar{z})=\phi_0-i(j_0\ln z+\bar{j}_0\ln \bar{z})+i\sum_{0\neq n\in\mathbb{Z}}\frac{1}{n}\lb j_n z^{-n}+\bar{j}_n\bar{z}^{-n}\rb\ .
\end{eqnarray}
The operator $\phi_0$ and $j_0$ are canonical conjugates of each other: $[\phi_0,j_0]=i$.
The $U(1)$ symmetry group is generated by $U_a=e^{i a j_0}$. 
The vertex operator $V_k(z,\bar{z})=:e^{ik \phi(z,\bar{z})}:$ acting on the vacuum creates eigenstates of the conjugate momenta $j_0\ket{k}=k\ket{k}$ with $\ket{k}=V_k(0)\ket{\Omega}$ and $\braket{k|k'}=\delta_{kk'}$. In fact, the vertex operator satisfies $[j_0,V_k]=k V_k$ which implies that it is a unitary intertwiner.

We can consider $\alpha(z)$ functions on the circle and the unitary vertex operator $V(\alpha)=:e^{i\phi(\alpha)}:$ with $\phi(\alpha)=\int dz \alpha(z)\phi(z)$. Under the transformation $\phi\to \phi+2\pi$ the vertex operator should be invariant therefore the charge $q_\alpha=\int dz \alpha(z)$ is quantized. 
When $q_\alpha=0$ the vertex operator $V(\alpha)$ is charge-neutral but when $q_\alpha=\int dz\:\alpha(z)\neq 0$ it is an intertwiner of charge $q$. The dual group is $\mathbb{Z}$ and is generated by charged vertex operators $V(k \alpha)$ for $k\in\mbZ$. 
As in the case of fermions, adding one intertwiner of unit charge adds all of them because $V(\alpha)V^\dagger(\beta)$ with $q_\alpha=q_\beta$ is a charge-neutral operator.
The action of the dual group on the invariant algebra at point is
\begin{eqnarray}
\rho_\alpha(J(z))\equiv V(\alpha) J(z)V^\dagger(\alpha)=J(z)+\alpha(z)
\end{eqnarray}
which does not leave the neutral operators invariant. Instead it shifts it by an element of the center of the algebra \cite{buchholz1988current}.
The action of the dual group on the symmetry generator is
\begin{eqnarray}
\rho_\alpha(U_a)=V(\alpha) U_a V^\dagger(\alpha)=e^{i a q_\alpha} U_a\ .
\end{eqnarray}
The dual group is not compact, but we can formally define an average over the charged sector as a distribution
\begin{eqnarray}
\sum_{k=-\infty}^\infty \rho_{k\alpha}(U_a)=\frac{2\pi}{|q_\alpha|}\delta(a)\ .
\end{eqnarray}

\subsection{Example 5: Permutation group}

The simplest example of a non-Abelian group is the permutation group $S_3$. 
Consider three qubits and the symmetry group $S_3$ that swaps the qubits. The elements of the group are the identity, the two-cycles and the three cycles. The two-cycles are represented by $U_{(12)}=S_{12}$, $U_{(13)}=S_{13}$ and $U_{(23)}=S_{23}$ where $S_{ij}$ is the swap operator of site $i$ and $j$: $S_{(12)}=\sum_{ab}\ket{ab}\bra{ba}$. The three-cycles are $U_{(123)}=\sum_{abc}\ket{abc}\bra{bca}$ and $U_{(132)}=\sum_{abc}\ket{abc}\bra{cab}$. 
The invariant algebra $\mA$ is the set of $4\times 4$ dimensional matrices $\ket{\alpha_i}\bra{\alpha_j}$ where $\ket{\alpha_i}$ are invariant vectors of $S_3$: $\ket{\alpha_0}=\ket{000}$, $\ket{\alpha_1}=\ket{111}$,
 $\ket{\alpha_2}=\frac{1}{3}(\ket{001}+\ket{010}+\ket{100})$ and $\ket{\alpha_3}=\frac{1}{3}(\ket{011}+\ket{101}+\ket{110})$. 

The Hilbert space has two sectors $\mH=(\mK_1\otimes \mH_1)\oplus (\mK_2\otimes \mH_2)$. 
The vacuum representation $\mK_1$ is the trivial one-dimensional representation, and $\mH_1=\overline{\mA\ket{000}}$ is the Hilbert space of states invariant under $S_3$ that is four dimensional and spanned by $\ket{\alpha_i}$. The Hilbert space $\mK_2$ is the two-dimensional irreducible representation of $S_3$ corresponding to  the Young tableaux \ytableausetup{smalltableaux}
$\ydiagram{2,1}$. 
The vectors $\ket{v_0}=\ket{100}-\ket{001}$, and $\ket{v_1}=2\ket{010}-(\ket{100}+\ket{001})$ provide a basis for this representation. It is straightforward to see that the action of $S_3$ leaves the two-dimensional subspace spanned by these vectors invariant.
Acting with the invariant algebra, in particular $\ket{\alpha_3}\bra{\alpha_2}$ on these vectors generates two perpendicular vectors $\ket{v_2}=\ket{011}-\ket{110}$ and $\ket{v_3}=2\ket{101}-(\ket{011}+\ket{110})$, and the new two-dimensional subspace is also preserved under the action of $S_3$. 
The sector $\mK_2\otimes \mH_2$ is the four dimensional subspace $\overline{\mA\ket{v_1}}$.
There is no totally anti-symmetric representation for qubits.

\subsection{Example 6: The $O(N)$ model}

Consider a real vector field $\Phi(f)$ with $N$ components of form $\varphi^{(j)}(f_j)$ and $f$ a collection of functions $f_1,\cdots ,f_N$. The algebra $\mF$ is generated by the Weyl operators $W(f)=e^{i \sum_j \varphi^{(j)}(f_j)}$. 
The symmetry group $O(N)$ acts on the vector fields which is equivalent to rotating $f_i$: $U_g W(f)U_g^\dagger=W(g.f)$ and $(g.f)_i=\sum_j g_{ij} f_j$. The invariant algebra $\mA$ is the algebra of $O(N)$ singlets generated by operators like $\Phi(f)\cdot \Phi(f)=\sum_i \varphi^{(i)}(f_i)\varphi^{(i)}(f_i)$. The vacuum sector is $\overline{\mA\ket{\Omega}}$.
The other sectors correspond to other irreducible representations of $O(N)$. Take the operator 
$\Phi(T)=\sum_{i_1,\cdots, i_k=1}^N T^{i_1,\cdots, i_k}\varphi^{(i_1)}(f_1)\cdots \varphi^{(i_k)}(f_k)$ where the tensor $T$ has symmetries under the permutation of indices that is characterized by a young tableaux $\lambda=(\lambda_1,\cdots,\lambda_s)$ with the total number of boxes $k=\sum_{j=1}^s \lambda_i$. Such operators acting on the vacuum sector take us to the charged sector with the irreducible representation characterized by the Young tableaux $\lambda$ and dimension $\dim(\lambda)$. One can find an orthonormal basis of such operators $\Phi(T_j)$ with $j=1,\cdots ,\dim(\lambda)$ \cite{buchholz1988current}.

\section{Bi-local intertwiners}\label{sec:4}

\begin{figure}[h]
 \centering
 \begin{subfigure}{.3\textwidth}
   \centering
   \includegraphics[width=.8\linewidth]{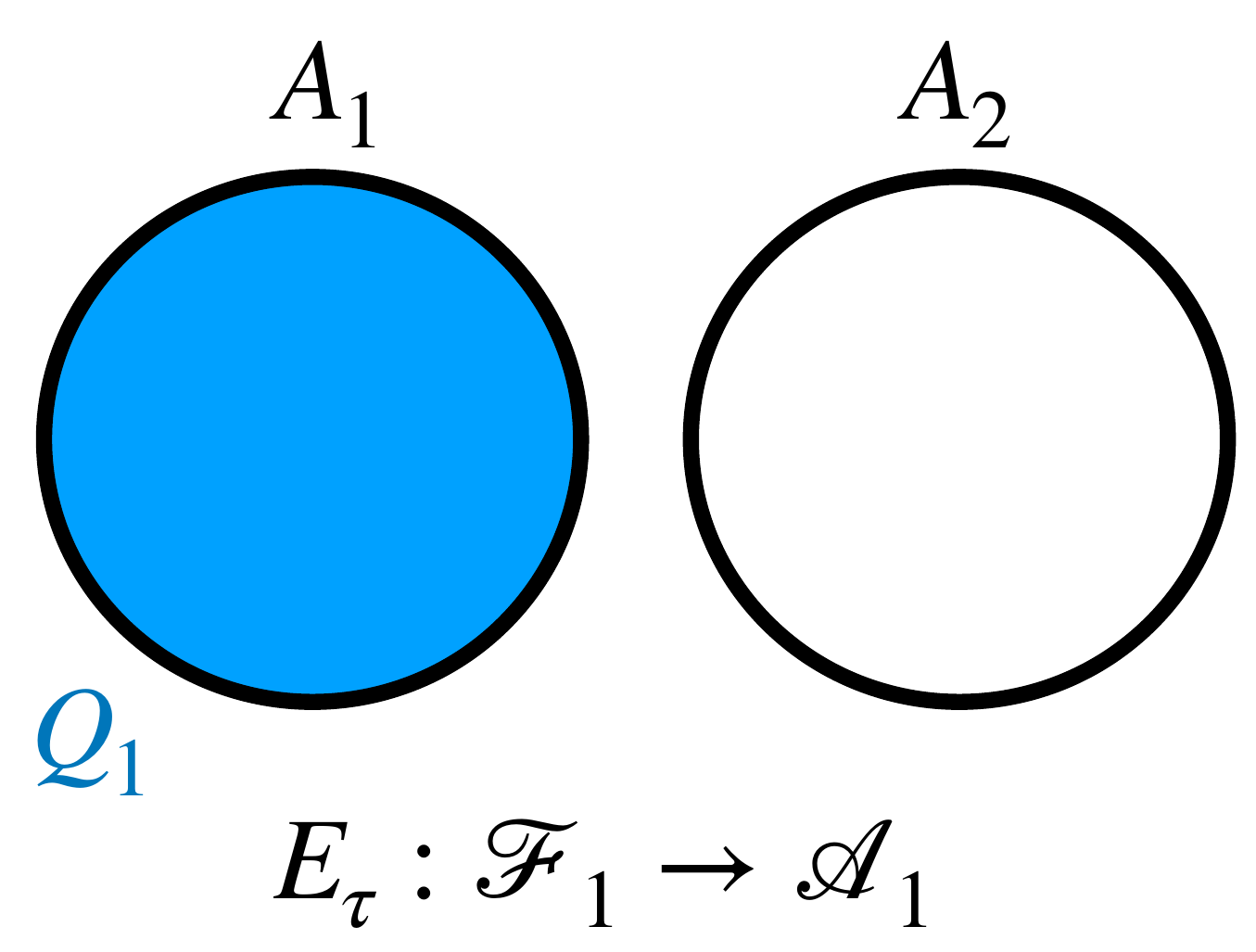}
   \caption{}
   \label{fig5:sub1}
 \end{subfigure}
 \begin{subfigure}{.3\textwidth}
   \centering
   \includegraphics[width=.8\linewidth]{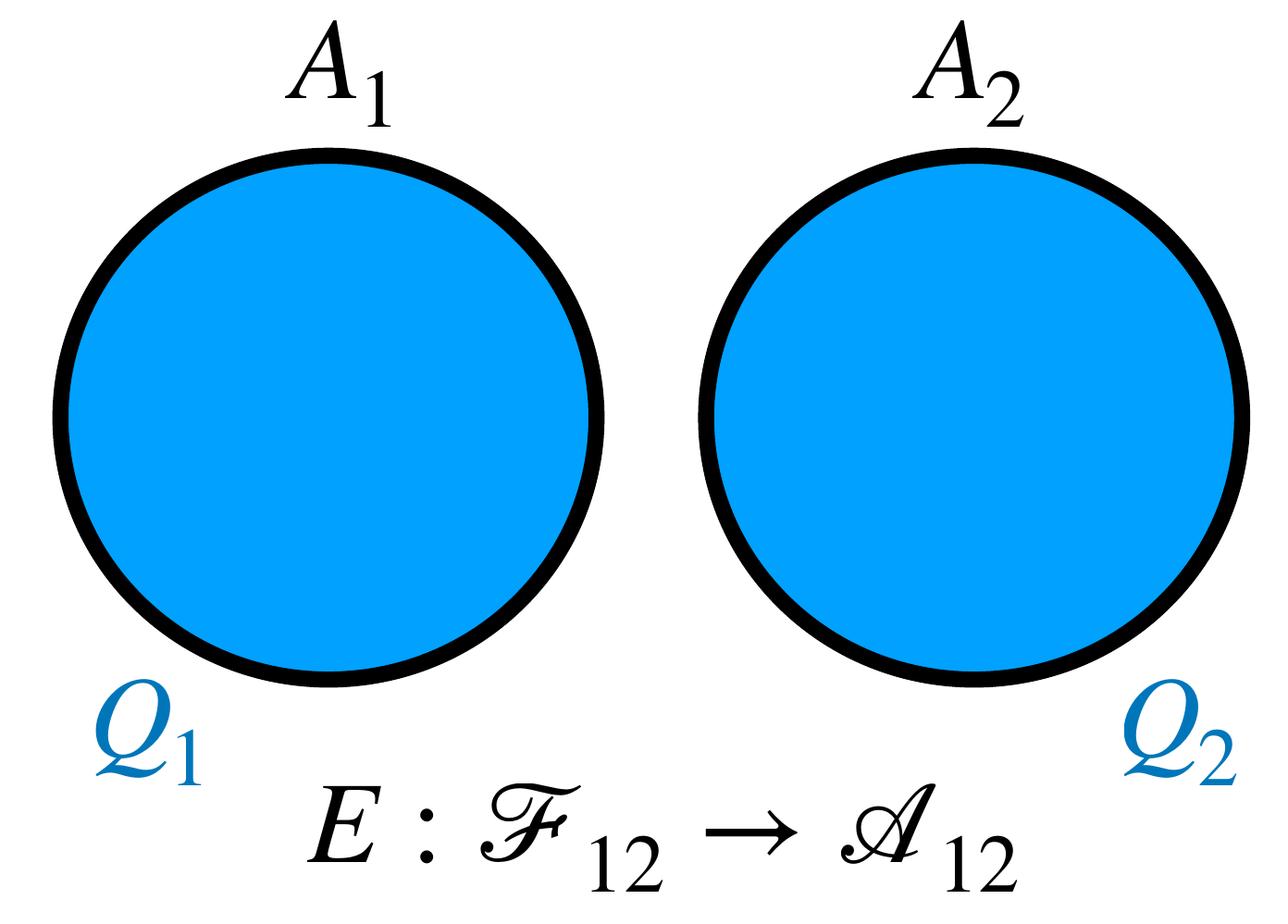}
   \caption{}
   \label{fig5:sub2}
 \end{subfigure}
 \begin{subfigure}{.3\textwidth}
   \centering
   \includegraphics[width=.8\linewidth]{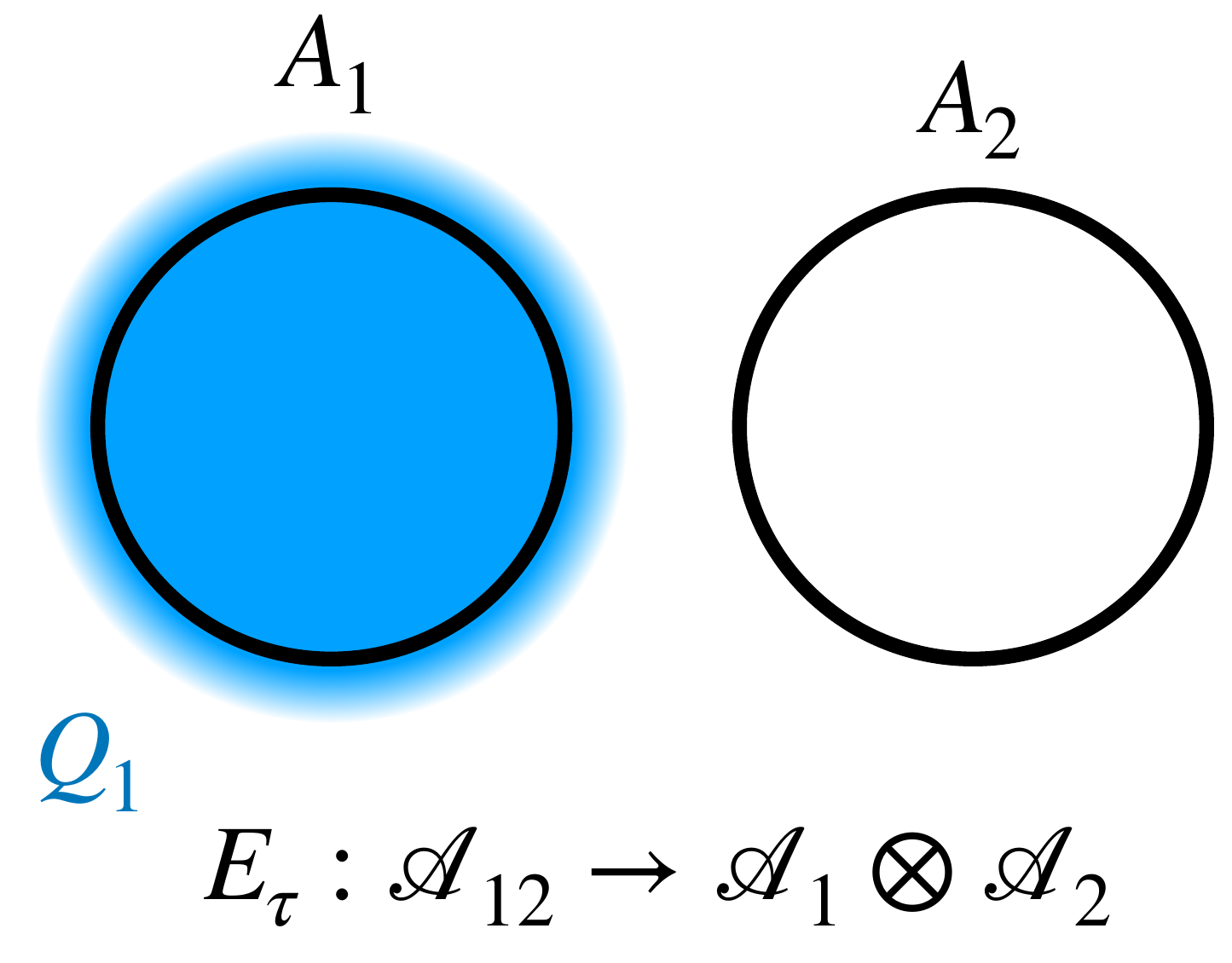}
   \caption{}
   \label{fig5:sub3}
 \end{subfigure}
 \caption{\small{Consider the operator $e^{i g \int_{x\in B} c(x) j(x)}$ where $j(x)$ is the charge density and the region $B$ is the blue region. (a) On a lattice we pick $c(x)=1$ that is the twist operator $\tau_g=e^{i g Q_1}$. It generates the action of the symmetry group on the local algebra of $A_1$. Averaging over $\tau_g$ is a conditional expectation that projects $\mF_1$ to $\mA_1$. (b) The action of the symmetry on the region $A_{12}$ is given by $e^{i g (Q_1+Q_2)}$. Averaging over this unitary projects from $\mF_{12}$ to $\mA_{12}$ c) In a QFT choosing $c(x)=1$ in $A_1$ and $c(x)=0$ outside of $A_1$ leads to an operator that has a violent behavior at the boundary of $A_1$ due to the discontinuity in $c(x)$. If there is a gap between $A_1$ and $A_2$ we can choose a $c(x)=1$ inside $A_1$ and make it smoothly fall out to zero without entering region $A_2$. This is the analog of the twist operator in a QFT. Averaging over this twist projects from $\mA_{12}$ down to $\mA_1\otimes \mA_2$.}}
 \label{fig5}
\end{figure}

Consider a multi-partite quantum system on a lattice with a symmetry $U_g=e^{i g Q}$ and local algebras $\mF_A$ associated with each region $A$ (collection of sites on a lattice or a region of space).\footnote{For the sake of the argument we have assumed  $G$ is a Lie group. However, the discussion applies to any group $G$.} 
We say a symmetry of the global algebra $\mF=\mF_{AA'}$ is internal if it preserves local algebras: 
\begin{eqnarray}\label{localaction}
 U_g a U_g^\dagger\in \mF_A\qquad\forall a\in \mF_A\ .
 \end{eqnarray}
There is a unitary group $\tau_g=e^{i g Q_A}$ localized in $A$ that generates the group action in (\ref{localaction}) for operators in $\mF_A$; see figure \ref{fig5}. In section \ref{sec:2} we called the operator $\tau_g$ the twist and its corresponding group the twist group $G_\tau$. When the group is Abelian $\tau_g$ is charge-neutral $U_h\tau_g U_h^\dagger=U_g$ and provides a center for the algebra of neutral operators. When the group $G$ is non-Abelian the operator $P_r$ in (\ref{Pr}) is in the center of the algebra: $Z=\oplus_r \lambda_r P_r$. 

Locality implies that $\mF_A$ commutes with the algebra of the complementary region $\mF_{A'}$. Define the commutant of algebra $\mF_A$ to be $\mF'_A$: the set of all operators in the global algebra $\mF_{AA'}$ that commute with $\mF_A$. From locality it follows that $\mF_{A'}\subset \mF'_A$. We say the region $A$ has the {\it duality} property if $\mF'_A=\mF_{A'}$. The full algebra of all charged operators satisfy the duality property, however the algebra of charge-neutral operators $\mA$ violate it. For instance, on a lattice the total charge is $Q=Q_A+Q_{A'}$ and $\mH_{AA'}=\mH_A\otimes \mH_{A'}$ the action of the symmetry transformation on $\mA_A$ is captured by the twist operator $\tau_g=e^{i g Q_A}$. The local algebra $\mA_A$ has a non-trivial center  $\mathbb{Z}_A=\oplus_r \lambda_r \mathbb{I}_r$ with $r$ irreducible representations of $\tau_g$ and $\lambda_r$ complex numbers. The duality relation for charge-neutral algebras is: 
$\mA'_A=\mathbb{Z}_A\otimes \mA_{A'}$.  
Note that here the commutant $\mA'_A$ is defined to be the algebra of operators in $\mA_{AA'}$ that commute with $\mA_A$. On a lattice, the failure of duality is due to a non-trivial center for the algebra of charge-neutral operators. However, in QFT the duality property can fail even though the local charge-neutral algebra has a trivial center. The reason is that the operator $P_r$ defined in (\ref{Pr}) is not part of the local algebra of region $A$ because it acts singularly on the boundary of $A$.

More generally, consider the region $A_{12}=A_1\cup A_2$ with two disconnected pieces $A_1$ and $A_2$. On a lattice the algebra of all charged particles is {\it additive} that is to say $\mF_{12}=\mF_1\otimes \mF_2$. In QFT, the additivity property holds when $A_1$ and $A_2$ are not touching.\footnote{We have assumed that QFT has the split property \cite{haag2012local}.} Both on a lattice or in QFT when we restrict to the subalgebra of locally charge-neutral operators additivity fails: $\mA_1\otimes \mA_2\neq \mA_{12}$. Of course, $\mA_1\otimes\mA_2$ is a subalgebra of $\mA_{12}$ but there exist operators in $\mA_{12}$, namely the bi-local intertwiners, that are not generated in $\mA_1\otimes \mA_2$. The bi-local intertwiner adds a charge $q$ to region $A_1$ and the opposite charge $-q$ to the region $A_2$ so that the total charge $Q_1+Q_2$ is conserved. The action of the symmetry group on $\mF_1$ can be captured by a local transformation $e^{i g Q_1}$ on a lattice. In QFT, the operator $e^{i g Q_1}$ has a singular behavior at the boundary of $A_1$. However, as long as there is a gap between region $A_1$ and $A_2$ there is a unitary transformation $\tau_g$  that matches $e^{i g Q_1}$ on $A_1$ and has a smooth tail that leaks outside of $A_1$ but does not enter $A_2$; see figure \ref{fig5}.  In analogy with the lattice systems, we call this operator the twist and the symmetry it generates the twist group:
\begin{eqnarray}
&&b_1\to e^{i g Q}b_1 e^{-i g Q}=\tau_g^\dagger b_1 \tau_g\qquad \forall b_1\in\mF_1\nn\\
&&[\tau_g,a_2]=0,
\qquad \forall a_2\in \mA_2\ .
\end{eqnarray}
In QFT, the local neutral-algebra has a trivial center.
When $G$ is compact one has the conditional expectation $E:\mA_{12}\to \mA_1\otimes \mA_2$ that is an average over the twist group:
\begin{eqnarray}
E(b)=\frac{1}{|G|}\int_{g\in G} dg\: \tau_g^\dagger b \tau_g\ .
\end{eqnarray}
By construction, the conditional expectation above sets any operators charged under $Q_1$ including bi-local intertwiners to zero.
The conditional expectation projects down to the invariant algebra. To go in the opposite direction, we need to enlarge the algebra $\mA_1\otimes \mA_2$ by adding the bi-local intertwiners to obtain $\mA_{12}$. Enlarging an algebra $\mA$ by the intertwiners of symmetry $G$ is mathematically described by the crossed-product of the algebra with its dual group, $\mA_{12}=(\mA_1\otimes \mA_2)\rtimes \hat{G}$; see appendix \ref{app:A} for details.

In QFT, there is no local Hilbert space $\mH_1$, and we only have the global Hilbert space $\mH$ and local algebras $\mA_1$.
In a QFT with charges, analogously, we have the global Hilbert space of type $\oplus_r \mK_r\otimes \mH_r$. The intertwiner $\ket{r,i}\bra{0}\otimes 1$ takes us from the global vacuum to the global charged sector $\ket{r,i}$ but it might not be localized in region $A$. We come back to this issue in section \ref{sec:5.3}. Similar to the Abelian case where we added $\ket{q}\bra{q+1}$ to get the unitary $\hat{U}_1$, we would like to extend the domain of $\ket{r,i}\bra{0}$ to an operator that adds charge $r$ to any state. The tensor product of two irreducible representations $r$ and $r'$ is a direct sum of irreducible representation with Clebsch-Gordon coefficients. 
A charged operator that is localized in $A$ commutes with all $a'\in\mA'$ and removes a charge $r$ \cite{doplicher1969fields,doplicher1970fields,araki1999mathematical}:
 \begin{eqnarray}
 &&V_{r,i}(\ket{r,i}\otimes a\ket{\Omega})=a\ket{\Omega}\nn\\
 &&V_{r,i}a\ket{\Omega}=\ket{r^*,i}\otimes a \ket{\Omega},
 \end{eqnarray} 
where $r^*$ is the conjugate representation of $r$. The action of $V_{r,i}$ on a vector $\ket{r',j}\otimes \ket{\Omega}$ is decided by the Clebsch-Gordon coefficients in the tensor product of representations $r$ and $r'$.
 The dual transformation maps the algebra of charge-neutral operators $\mathbb{I}_r\otimes a$ back to the vacuum sector 
 \begin{eqnarray}\label{dualmap}
  \rho_r(a)=\sum_i V_{r,i} a V_{r,i}^\dagger\ .
 \end{eqnarray}
The map $\rho_r(a)$ maps the charge-neutral operators $\mA$ to itself and since it is the representation of the local algebra it respects the multiplication rule\footnote{Such a map is called an endormophism of the algebra.}
 \begin{eqnarray}
  \rho_r(a_1a_2)=\rho_r(a_1)\rho_r(a_2)\ .
 \end{eqnarray}
 The condition above together with (\ref{dualmap}) imply that $V_{r,i}$ should satisfy the algebra
 \begin{eqnarray}
 &&V_{r,i}^\dagger V_{r,j}=\delta_{ij}\nn\\ 
 && \sum_i V_{r,i}V_{r,i}^\dagger=1\ .
 \end{eqnarray}
 The algebra above is called the Cuntz algebra \cite{doplicher1990there}. 
The Cuntz algebra has no finite dimensional representations; however, it is easy to build representations of the Cuntz algebra in infinite dimensions. For instance, take the Hilbert space of a particle on a circle and split it into two sectors defined by projections to the even and odd momenta $P_+=\sum_{k}\ket{2k}\bra{2k}$ and $P_-=\sum_k \ket{2k+1}\bra{2k+1}$. The isometries $V_1=\sum_k \ket{2k+1}\bra{k}$ and $V_2=\sum_k \ket{2k}\bra{k}$ satisfy the Cuntz algebra with $i=1,2$.
 
 The particle number is not conserved in relativistic QFT. Acting with $V_{r,i}^\dagger$ creates one charged particle  but applying it again we can have several charged particles. There is a subalgebra of the Cuntz algebra that corresponds to a sector with one charged particle $\sum_{r,i}a_{ij} V_{r,i}V_{r,j}^\dagger$, where $a_{ij}$ are invariant operators. These operators can be represented by a $d_r\times d_r$ matrix algebra.
 
The operators $V_{r,i}$ satisfies the non-Abelian intertwiner equation
\begin{eqnarray}
 V_{r,i}a=\rho_r(a) V_{r,i},\forall a\in\mA
\end{eqnarray}
and $V_{r,i}^\dagger$ acting on the vacuum sector creates charged states in representation $r$: $\ket{r,i}=\sqrt{d_r}V_{r,i}^\dagger\ket{\Omega}$. The factor $\sqrt{d_r}$ is needed to make sure $\braket{r,i|r,i}=1$. There are also the states in the conjugate representation that are created by $\ket{r^*,i}=V_{r,i}\ket{\Omega}$.\footnote{ Note that in this case there is no need for a factor $\sqrt{d_r}$ to normalize the state.} The conjugate representation is
\begin{eqnarray}
\rho_{r^*}(a)=\frac{1}{d_r}\sum_i V_{r,i}^\dagger a V_{r,i}\ .
\end{eqnarray}
In a charged sector the expectation value of a charge neutral operator satisfies
\begin{eqnarray}
\sum_i\braket{r,i|a|r,i}=\braket{\Omega|\rho_r(a)|\Omega}\ .
\end{eqnarray}
If $\rho_r(a)=a$, one cannot distinguish charged sectors. However, if $\rho_r(a)\neq a$, this is no longer true. An example of this is the compact boson example:
\begin{equation}
    \rho_\alpha(J(z))=J(z)+\alpha(z)\ .
\end{equation}

The group transformation $U_g=\oplus_r U^r_g$ acts on the intertwiner according to the equation 
\begin{eqnarray}\label{repMat}
U_g^\dagger V_{r,i}=D_r(g)_{ij} V_{r,j}U_g^\dagger
\end{eqnarray}
where $D_r(g)_{ij}$ are the matrix elements of the representation matrix $D_r(g)$ with the orthogonality relations\footnote{In the case of an Abelian group this is $U_g^\dagger V_q=e^{-\frac{2\pi i g q}{|G|}} V_q U_g$.}
\begin{eqnarray}\label{ortho}
\frac{d_r}{|G|}\sum_g D_r(g)_{ik} D_{r'}(g)_{jl}^*=\delta_{rr'}\delta_{ij}\delta_{kl}\ .
\end{eqnarray}

In QFT, in analogy with lattice systems, it is tempting to take the local algebra of $A$ to be all charged operators $\ket{r,i}\bra{r',i'}\otimes a$, however as we discussed above, in QFT $\ket{r,i}\bra{0}\otimes 1$ is not localized in $A$, and the charge neutral algebra has no non-trivial center. 
Local charges on $A$ are created by the intertwiners $V_{r,i}^\dagger$, instead of $\ket{r,i}\bra{0}\otimes 1$. 
Therefore, we define the local algebra of charge operators to be the algebra generated by charge neutral operators $a$ and the isometries $V_{r,i}^\dagger$.
Consider charged operators $\sum_i a_i V_{r,i}$. 
Bi-local intertwiners create/annihilate a charge in $A_1$ and create/annihilate the opposite charge in $A_2$ so that the net charge is preserved:
\begin{eqnarray}
\mathcal{I}^{(r)}_{12}=\sum_i (V^{(1)}_{r,i})^\dagger V^{(2)}_{r,i}
\end{eqnarray}
with $V^{(1)}_{r,i}$ and $V^{(2)}_{r,i}$ supported on $A_1$ and $A_2$, respectively.
This is a unitary map in the global algebra that is charge-neural. However, from the point of view of algebra $\mA_1$ it is an intertwiner.

In the remainder of this section, we provide several examples of bi-local intertwiners in finite quantum systems and QFTs.

\subsection{Example 1: Qubits}

Consider two qubits in $\mH_1\otimes \mH_2$ and the symmetry group $\mathbb{Z}_2$ corresponding to the action of $(-1)^Q=\sigma_z\otimes \sigma_z$ where the total charge $Q=Q_1\otimes \mathbb{I}+\mathbb{I}\otimes Q_2$ and $Q_i=\frac{1}{2}(\mathbb{I}-\sigma_z)$ counts the number of excitations ``$\ket{1}$''. The action of the symmetry group on the local algebra $\mF_1$ is captured by the twist group $\tau_1=(-1)^{Q_1}$ that is localized in $A_1$.
The algebra of global charge-neutral operators $\mA_{12}$ is the set of all operators that commute with $Q$. The charge neutral sub-algebras $\mA_1\otimes \mathbb{I}_2$ and $\mathbb{I}_1\otimes \mA_2$ commute with $Q$, however, $\mA_{12}$ includes more operators. In particular, the operator that creates a charge on site one and annihilates it on site two commutes with $Q$: \begin{eqnarray}
[Q,\sigma^\pm\otimes \sigma^\mp]=0,\qquad \sigma^\pm=\frac{1}{2}(\sigma^{(x)}\mp i\sigma^{y})\ .
\end{eqnarray}
The algebra $\mA_{12}$ also includes operators that increase $Q$ by two units, $\ket{00}\bra{11}\in \mA_{12}$ and its $\dagger$.
The subspace $\mH'_1=\mA_{12}\ket{00}$ that is spanned by $\ket{00}$ and $\ket{11}$ is the zero charge sector and the charged sector is $\mH'_2=\mA_{12}\ket{01}$ which is spanned by $\ket{01}$ and $\ket{10}$.

The subalgebra of $\mA_{12}$ invariant under the twist group $(-1)^{Q_1}$ is $\mA_1\otimes \mA_2$.
Each sector $\mH'_1$ and $\mH'_2$ further splits into two sectors depending on the eigenvalue of $\sigma_z^{(1)}$. 
The operator $\ket{11}\bra{00}$ is an internal intertwiner for the twist group that is a unitary in $\mH'_1$, and $\sigma^-\otimes\sigma^+=\ket{10}\bra{01}$ is an internal intertwiner for the twist group in $\mH'_2$. Local intertwiners  create a pair of charge/anti-charge excitations.
 The group average over the twist is a conditional expectation $E_\tau:\mA_{12}\to \mA_1\otimes \mA_2$ that washes out local intertwiners:
\begin{eqnarray}
E_\tau(b)=\frac{1}{2}\lb b+\sigma^{(1)}_z b\sigma^{(1)}_z\rb .
\end{eqnarray}

This can be easily extended to $n$ qubits with the global symmetry $\mathbb{Z}_2$ that is measured by the total charge $(-1)^Q=\otimes_{i=1}^n\sigma_z^{(i)}$ and the local charge associated with the region $A$ that is the first $m$ qubits $\tau=(-1)^{Q'}=\otimes_{i=1}^m\sigma^{(i)}_z$. The global Hilbert space splits into two sectors $\mH=\mH_+\oplus \mH_-$ where $\mH_\pm$ is spanned by all $\ket{s_1,\cdots s_n}$ with $s_1s_2\cdots s_n=\pm 1$. The twist symmetry $\tau=(-1)^{Q'}$ further splits each sector into two: $s_1s_2\cdots s_m=\pm 1$. The operator $\ket{s_1\cdots s_m,t_{m+1}\cdots t_n}\bra{s'_1\cdots s'_m,t'_{m+1}\cdots t'_{n}}$ with $s_1\cdots s_m=-1=-s'_1\cdots s'_m$ and $t_{m+1}\cdots t_n=\pm 1=\mp (t'_{m+1}\cdots t'_n)$ is an example of a local intertwiner.

As an example of a region with two non-overlapping pieces consider the local algebras $\mA_{12}$ and $\mA_1\otimes \mA_2$ where there are a total of three qubits. We first check the duality property. 
Once we include the centers of local algebra the duality property holds: $\mA'_{12}=\mathbb{Z}_{12}\otimes \mA_3$ and $\mA'_3=\mathbb{Z}_3\otimes \mA_{12}$. Note that the operator 
 $\tau_{13}=(-1)^{Q_1+Q_3}=\sigma_z^{(1)}\otimes\sigma_z^{(3)}$ is in $\mA'_3$ but not in $\mA_{12}$. In fact, if we only add $\tau_{13}$ to $\mA_{12}$ we generate the full $\mathbb{Z}_3\otimes \mA_{12}$.
 The operator $\tau_{13}$ is a twist operator similar to the ones in QFT because it acts on $\mA_1$ like $(-1)^{Q_1}$, it is supported outside of $A_1$ but its support does not enter $A_2$. We learn that another way to express the duality relation for charge-neutral algebras is by enlarging $\mA_{12}$ with the twist $\tau_{13}$. In mathematical language, we write the crossed product $\mA'_3=\mA_{12}\rtimes G_{13}$ where $G_{13}$ is the symmetry group generated by $\tau_{13}$.
We could replace $\tau_{13}$ with $\tau_{23}$ or $\tau_3$ and the result remains the same.
However, for the opposite region we have to enlarge $\mA_3$ by $\tau_{12}$ to obtain $\mA'_{12}$: $\mA'_{12}=\mA_3\rtimes G_{12}$.
If we have four qubits, then the equations become more symmetric:
\begin{eqnarray}\label{fourqubits}
&&\mA'_{12}=\mA_{34}\rtimes G_{12}\nn\\
&&\mA'_{34}=\mA_{12}\rtimes G_{34}\ .
\end{eqnarray}
A much simpler way to write the duality equation for charge-neutral algebras is $\mA'_A=\mA_{A'}\rtimes G$ where $G=(-1)^Q$ is the generator of the symmetry in the global algebra.

The interplay between duality and additivity of local algebras plays an important role in the study of quantum systems with symmetries \cite{Casini:2019kex}. 
The action of a symmetry on a local region $A_i$ is captured by the twist group $G_{ik}$ generated by $\tau_{ik}$ with $A_k$ some region outside of $A_i$. On a lattice, one can take the twist to be $\tau_i$. 
Denote the local intertwiner that creates a charge on $A_i$ and annihilates it in $A_j$ by $\mathcal{I}_{ij}$. It generates a group dual to the twist group $G_i$; or $G_{ik}$ for $k\neq j$ in QFT. When the algebra is Abelian this duality transformation is a Fourier transform and indeed we find 
$[\mathcal{I}_{ij},\tau_{i}]\neq 0$. In the qubit example, we have $[\mathcal{I}_{ij},\tau_i]=2\mathcal{I}_{ij}$.

\subsection{Example 2: Free relativistic fermions}

Consider free fermions in $(1+1)$-dimensions. As we discussed in section \ref{sec:3.4} the symmetry of the global algebra is $(-1)^N$ where $N$ is the total number of fermions, and the invariant global algebra is all operators with an even number of fermions. The local algebra of a region is generated by $\Psi(f_A)$ with $f_A$ any bounded complex function supported only a region $A$.\footnote{We thank Edward Witten for pointing this out to us.} The symmetry acts on the local algebra as $\tau=(-1)^{N_A}$ where $N_A$ is the total number of fermions in a region $A$. This operator is discontinuous at the boundary of $A$ and we can smooth it outside of $A$. The Hilbert space splits into four sectors corresponding to two charges $(N \mod 2)=0,1$ and $(N_A \mod 2)=0,1$. The operator $\Psi(f_{A'})\Psi^\dagger(g_A)$ creates a pair of charge/anti-charge particles in $A$ and $A'$. It is a bi-local intertwiner for $\mA_A$.

If we take two regions of space $A_1$ and $A_2$ that are non-overlapping and non-touching the complementary region also has two disconnected pieces. This is analogous to the case of four qubits we discussed above.\footnote{In higher than $(1+1)$-dimensions the complement of $A_{12}$ is connected and the three qubit example is a better analogy.} In addition to $\mA_1\otimes \mA_2$ the algebra of $\mA_{12}$ includes the intertwiners from region $A_1$ to $A_2$ that are $\Psi^\dagger(f_1)\Psi(f_2)$ with $f_i$ supported in $A_i$. The twist operator $\tau=(-1)^{Q_1}$ needs to be smoothed out outside of $A_1$ without leaking inside $A_2$. We call the smooth twist operator $\tau_{13}$ because it is supported on $A_{13}$ and acts like $(-1)^{N_1}$ on $\mA_1$.  
The group average over $\tau_{13}$ is a conditional expectation $E_\tau:\mA_{12}\to \mA_1\otimes \mA_2$:
\begin{eqnarray}
E_\tau(b)=\frac{1}{2}\lb b+\tau_{13}^{-1} b\tau_{13}\rb.
\end{eqnarray}
It kills the local intertwiners: $E(\Psi(f_1)\Psi^\dagger(f_2))=0$.
In QFT, there are no local density matrices, instead the local state is a restriction of the global pure state to the local algebra:
\begin{eqnarray}
\omega(a_1\otimes a_2)=\braket{\Omega|a_1\otimes a_2|\Omega}.
\end{eqnarray}
The invariant state is 
\begin{eqnarray}
(E_\tau^*(\omega))(a_1\otimes a_2)=\frac{1}{2}\lb \omega(a_1\otimes a_2)+\omega(\tau_{13}^{-1}(a_1\otimes a_2)\tau_{13})\rb
\end{eqnarray}
which can be thought of as the restriction of the global density matrix 
\begin{eqnarray}
\frac{1}{2}\lb \ket{\Omega}\bra{\Omega}+\tau_{13}\ket{\Omega}\bra{\Omega}\tau_{13}^{-1}\rb
\end{eqnarray}
to the local algebra $\mA_{12}$. 

\subsection{Example 3: $U(1)$ current algebra}

\begin{figure}[h]
 \centering
 \begin{subfigure}{.4\textwidth}
   \centering
   \includegraphics[width=1.0\linewidth]{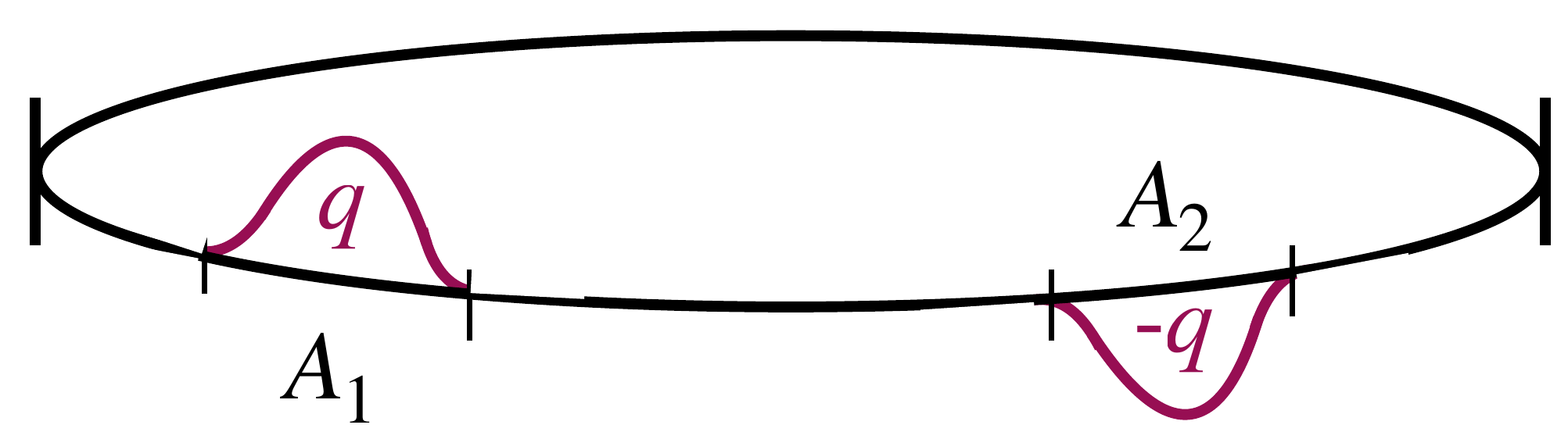}
   \caption{}
   \label{fig6:sub1}
 \end{subfigure}
 \begin{subfigure}{.4\textwidth}
   \centering
   \includegraphics[width=1.0\linewidth]{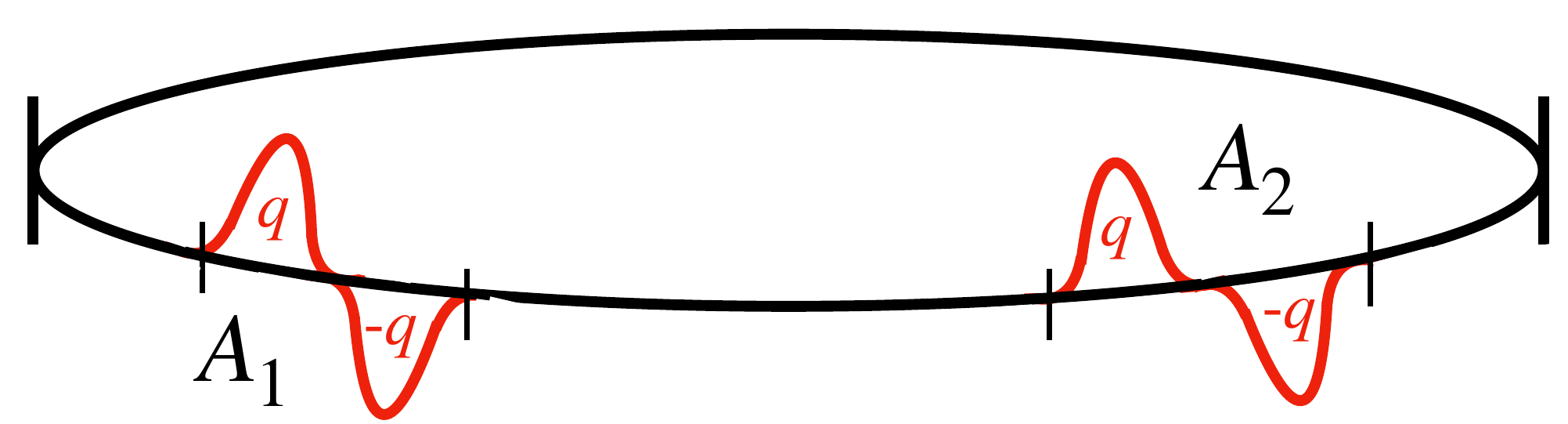}
   \caption{}
   \label{fig6:sub2}
 \end{subfigure}
 \caption{\small{(a) The bi-local intertwiners in $A_1\cup A_2$ conserve the total charge, $\int_{A_1\cup A_2} dz \:\alpha(z)=q=0$. (b) The subalgebra $\mathcal{N}_1\subset \mA_1$ does not have any operators that create and annihilate charges inside $A_1$ (the red excitations). Such an algebra is generated by $J(f)$ with functions localized in $A_1$.}}
 \label{fig6}
\end{figure}

In the free $(1+1)$-dimensional compact boson model, the symmetry group is $e^{i a j_0}$ and the Hilbert space has many sectors $\ket{\alpha}$ with the vertex operators $V(\alpha)=:e^{i \Phi(\alpha)}:$ with $\alpha$ some function on the circle intertwining them. If we consider the local algebra generated by $e^{i J(f_A)}$ with $f_A$ some smooth function supported only on $A$ then the total charge on $A$ is $j_0(A)=\frac{1}{2\pi}\oint_A J(z)$ where $A$ is some angle on the unit circle in radial quantization. The bi-local intertwiners between two non-touching, non overlapping regions $A_1$ and $A_2$ are $V(\alpha)$ with $\int_{A_1} dz \:\alpha(z)=q_A$ and $\int_{A_1\cup A_2} dz \:\alpha(z)=q=0$ so that they do not change the global sector; see figure \ref{fig6}.

\section{Intertwiners and Modular Theory}\label{sec:5}

In a Poincare-invariant QFT in $(d+1)$-dimensions, the global algebra of spacetime $\mF$ is generated by the bounded functions of the field operator $\Phi(f)$ with $\Phi(f)=\int d^{d+1}x f(x)\Phi(x)$ and $f(x)$ a solution to the classical equations of motion that respects the boundary conditions at infinity\footnote{Assumptions about the smoothness of the function $f$ are implicit in what is meant by a solution to the classical equations of motion.} \cite{haag2012local,araki1999mathematical,hollands2018entanglement}. This algebra is represented irreducibly on a global Hilbert space $\mH$. The local algebra $\mF_A\subset \mF$ is the subalgebra generated by $\Phi(f)$ where $f$ is only supported in $A$. The local algebra of QFT does not have an irreducible representation and there is no local Hilbert space \cite{witten2018aps}. The local algebra $\mF_A$ and that of the complementary region $\mF_{A'}$ both act on the global Hilbert space.
The local states are the restriction of the global state to the local algebra:
\begin{eqnarray}
 \omega_A(b)=\braket{\Omega|b\Omega},\qquad \forall b\in\mF_A\ .
\end{eqnarray}
Since there are no local Hilbert spaces there are no density matrices either. Modular theory is a mathematical formulation that allows us to define information theoretic quantities using only global states and local algebras, with no need for the existence of local density matrices; see \cite{hollands2018entanglement}. It applies to any quantum system from qubits to QFT. In QFT the algebras $\mF_A$ and $\mF_{A'}$ are isomorphic and the global vectors of QFT are analogous to the canonical purification of $\rho_A$ a density matrix of $A$ in terms of $\ket{\Omega}$ a pure state of a double copy Hilbert space $\mH_A\otimes \mH_{A'}$:
\begin{eqnarray}\label{canonicalform}
&&\omega=\sum_k p_k\ket{k}\bra{k}\nn\\
&&\ket{\Omega}=\sum_k \sqrt{p_k}\ket{k}_A\otimes\ket{k}_{A'}\ .
\end{eqnarray}

If $A_1$ and $A_2$ are two non-overlapping and non-touching regions of space, and $\mF_1$ and $\mF_2$ are their corresponding local algebras in QFT, the additive algebra of the union $A_{12}$ is the algebraic tensor product of local algebras $\mF_{12}=\mF_1\otimes \mF_2$.\footnote{In infinite dimensions, one has to be careful when tensoring von Neumann algebras since the weak closure of operators depends on the Hilbert space on which it is acting \cite{witten2018aps}. This is the so-called split property of QFT that we have assumed to hold in any reasonable model.}. There is no tensor product when the regions $A_1$ and $A_2$ touch. The algebra of invariant local operators $\mA$ has a trivial center because the twist operator $e^{i g Q_1}$ does not belong to $\mA_1$, however, when $A_1$ and $A_2$ are not touching the smoothed out twist commutes with both $\mA_1$ and $\mA_2$; see figure \ref{fig5}.

In section \ref{sec:2}, we argued that the correct entanglement measure in the presence of charges is the relative entropy in (\ref{final2}):
\begin{eqnarray}\label{relQFT}
S^{\mF_{12}}(E^*(\omega_{12})\|E^*(E_\tau^*(\omega_{12})))+S^{\mF_{12}}(E^*(E^*_\tau(\omega_{12}))\|E^*(E_\tau^*(\omega_1)\otimes \omega_{2}))
\end{eqnarray}
with the conditional expectations
\begin{eqnarray}
 &&E^*(\omega_{12})=\frac{1}{|G|}\sum_{g\in G}U_g\omega_{12} U_g^\dagger\nn\\
 &&E^*_\tau(\omega_{12})=\frac{1}{|G|}\sum_{g\in G}\tau_g\omega_{12} \tau_g^\dagger
\end{eqnarray}
where $U_g=e^{i g Q}$ and $\tau_g$ is the smoothed out $e^{i g Q_1}$. By a unitary rotation of local states we mean
\begin{eqnarray}
(U \omega U^\dagger)(b)=\omega(U^\dagger b U)\ .
\end{eqnarray}

We have structured this section in the following way:
In  section \ref{sec:5.1}, we start by a discussion of the local charged states and a lower and an upper bound on (\ref{relQFT}). In section \ref{sec:5.2}, we review the Tomita-Takesaki modular theory (see \cite{borchers2000revolutionizing} for a more detailed review) and compute the modular operators for charged states $\ket{r,i}$, and comment on the mirror operators in the presence of charges. Section \ref{sec:5.3} discusses the relation between the cocycle operator in modular theory and local charges. 
Finally, in section \ref{sec:5.4} we introduce a canonical enlarging of the algebra of QFT that decouples charged modes across the entangling surface.

\subsection{Charged states}\label{sec:5.1}

Consider the global invariant vector $\ket{\Omega}$ and its local state $\omega$ on region $A$.
Since $\ket{\Omega}=U_g\ket{\Omega}$, the expectation value of all charged operators of the form $b-E(b)$ vanishes in $\omega$:
\begin{eqnarray}
\omega(b)=\frac{1}{|G|}\sum_{g\in G}\braket{\Omega|U_g^\dagger b U_g\Omega}=\braket{\Omega|E(b)\Omega}\ .
\end{eqnarray}
All the charged states $\ket{r,i}$ are perpendicular to the vacuum since they belong to different superselection sectors. We denote by $\ket{r,i,A}=\sqrt{d_r}(V^{(A)}_{r,i})^\dagger\ket{\Omega}$ a state with a charge localized in region $A$.
A vector $\ket{\Phi}=\frac{1}{\sqrt{2}}(\ket{\Omega}+\ket{r,i,A})$ that superposes the vacuum with a charged state appears mixed to the local charge-neutral subalgebra of $A$:
\begin{eqnarray}
\braket{\Phi|a\Phi}=\frac{1}{2}(\omega(a)+\omega_{r,i}(a))
\end{eqnarray}
where $\omega_{r,i}(a)=\braket{r,i|a|r,i}$ is the local charged state, which turns out to be independent of $i$; see (\ref{indep}).
The same holds for the local state of the vector $\ket{\chi}=\frac{1}{\sqrt{2}}(\ket{r,i,A_1}+\ket{r,j,A_2})$. With respect to any charge neutral operator $a\in\mA_{12}$ the state seems mixed
\begin{eqnarray}
\braket{\chi|a\chi}=\frac{1}{2}(\omega_{r,i}(a)+\omega_{r,j}(a))\ .
\end{eqnarray}
This is because
\begin{eqnarray}
\braket{r,i,A_1|a|r,j,A_2}=d_r\braket{\Omega|V^{(1)}_{r,i} a (V_{r,j}^{(2)})^\dagger|\Omega}=\frac{d_r}{|G|}\sum_g\braket{\Omega|U_g^\dagger V_{r,i}^{(1)} a (V_{r,j}^{(2)})^\dagger U_g|\Omega}\ .
\end{eqnarray}
Using the transformation rule of the intertwiner in (\ref{repMat})
we find 
\begin{eqnarray}
&&\braket{r,i,A_1|a|r,j,A_2}=\frac{d_r}{|G|}\sum_{glk} D_r(g)_{ik}D_r(g)_{jl}^*\braket{\Omega|V_{r,k}^{(1)} a(V^{(2)}_{r,l})^\dagger\Omega}\nn\\
&&=\delta_{ij}\sum_k\braket{\Omega|V^{(1)}_{r,k}a(V^{(2)}_{r,k})^\dagger|\Omega}=\frac{\delta_{ij}}{d_r}\sum_k\braket{r,k,A_1|a|r,k,A_2}
\end{eqnarray}
where we have used (\ref{ortho}).
We learn that $\braket{r,i,A_1|r,j,A_2}\sim\delta_{ij}$ and when $i=j$ the expectation value of $a$ is independent of $j$:
\begin{eqnarray}\label{indep}
\omega_{r,j}(a)=\braket{r,j,A|a|r,j,A}=\sum_k\braket{\Omega|V_{r,k}aV_{r,k}^\dagger|\Omega}=\omega(\rho_r(a))\ .
\end{eqnarray}
Therefore, $\omega_r(a)\equiv \omega(\rho_r(a))=\omega_{r,j}(a)$ which implies that one cannot distinguish $\ket{r,i}$ and $\ket{r,j}$ using charge-neutral operators. 
For a general vector $\ket{\Psi}=\sum_{r,i}c_{r,i}\ket{r,i,A}$ 
we have
\begin{eqnarray}\label{purification}
&&(E^*(\psi))(a)=
\sum_{r,i}|c_{r,i}|^2 \braket{r,i,A|a|r,i,A}=\sum_r  \omega_r(a)\zeta_r=\psi(a),
\end{eqnarray}
where $\zeta_r=\sum_i |c_{r,i}|^2$.
 
Now, consider non-touching regions $A_1$ and $A_2$ and a global invariant state $U_g\ket{\Omega}=\ket{\Omega}$. The local states $\omega_{12}$ and $\omega_1\otimes \omega_2$ both have zero total charge $Q_1+Q_2=0$ and we only need to consider the charge neutral subalgebra $\mA_1\otimes \mA_2$ and the bi-local unitary intertwiners $\mathcal{I}_{12,r}=\sum_i (V^{(1)}_{r,i})^\dagger V_{r,i}^{(2)}$.
The bi-local intertwiner is a unitary operator that creates an entangled pair of charge/anti-charge particles $\mathcal{I}_{12,r}\ket{\Omega}=\ket{\mathcal{I}_{12,r}}$. These states are orthonormal: $\braket{\mathcal{I}_{r}|\mathcal{I}_{r'}}=\delta_{rr'}$ and $\braket{\mathcal{I}_{r'}|\mathcal{I}_s|\mathcal{I}_{r}}=\delta_{r's}\braket{\Omega|\mathcal{I}_r}$. 
They have an overlap with the vacuum state 
\begin{eqnarray}
\braket{\Omega|\mathcal{I}_{12,r}}=\sum_i \braket{\Omega|(V^{(1)}_{r,i})^\dagger V^{(2)}_{r,i}|\Omega}=\frac{1}{d_r}\sum_i \braket{r,i,A_2|r,i,A_1}\ .
\end{eqnarray}
The vacuum state has a non-zero amplitude to fluctuate to a state with multiple entangled pairs $\braket{\Omega|\mathcal{I}_{12,r_1}\cdots \mathcal{I}_{12,r_n}}\neq 0$. If the symmetry group is Abelian $\mathcal{I}_{12,r_1}\cdots \mathcal{I}_{12,r_n}=\mathcal{I}_{12,r_1+\cdots r_n}$. 

The average $E^*_\tau$ projects the algebra $\mA_{12}$ to $\mA_1\otimes \mA_2$ by discarding the bi-local intertwiner $\mcI_{12}$. The averaged state $E^*_\tau(\omega_{12})$ has zero amplitude for the creation of an entangled pair of charged particles between region $A_1$ and $A_2$. 
Adding any bi-local intertwiner $\mcI_{12}$ to $\mA_1\otimes \mA_2$ immediately enlarges it to $\mA_{12}$. If we want to isolate the contribution of any particular $\mcI_{12}=V^\dagger(x) V(y)$ with $x\in A_1$ and $y\in A_2$ to the relative entropy we need to find a subalgebra of $\mA_{12}$ that only includes this particular bi-local intertwiner, and none of the others.

There is a subalgebra of global charged neutral operators that has no bi-local intertwiners in it; that is to say we have discarded $\mcI_{12}$ for any non-touching $A_1$ and $A_2$. This is the algebra of QFT with no charge creation or annihilation operators.
For instance, in the example of the $U(1)$ current model, the algebra generated by $J(z)$ without any vertex operators is such a subalgebra. We denote such a subalgebra by $\mathcal{N}$. The restriction of $\mathcal{N}$ to a region $A_1$ gives a subalgebra $\mathcal{N}_{1}\subset \mathcal{A}_{1}$ and a conditional expectation that washes out any bi-local intertwiners within $A_1$.
The subalgebra $\mathcal{N}_1\otimes \mathcal{N}_2\subset A_{12}$ has no bi-local intertwiners within $A_1$, $A_2$ or in between $A_1$ and $A_2$.
Enlarging $\mathcal{N}_{1}\otimes\mathcal{N}_2$ by adding any $\mcI_{12}$ gives a subalgebra of $\mA_{12}$, rather than immediately generating the whole $\mA_{12}$. For instance, in regions $A_1$ and $A_2$ we can choose to add a bi-local intertwiner $V^\dagger(x) V (y)$ with $x\in A_1$ and $y\in A_2$; see figure \ref{fig6}. The relative entropy 
\begin{eqnarray}\label{lowerboundrel}
S^{(\mathcal{N}_1\otimes\mathcal{N}_2)\rtimes \mcI_{12}}(\omega_{12}\|E_\tau^*(\omega_{12}))\leq S^{\mA_{12}}(\omega_{12}\|E_\tau^*(\omega_{12}))
\end{eqnarray}
measures the contribution of this particular bi-local intertwiner, and we have used the monotonicity of relative entropy to get a lower bound on our entanglement measure due to bi-local intertwiners. 
The authors of \cite{Casini:2019kex} argued that the bi-local intertwiners with the minimal distance $|x-y|$ in between $A_1$ and $A_2$ give the tightest lower bound for the relative entropy $S(\omega_{12}\|E_\tau^*(\omega_{12}))$. In the literature, such bi-local intertwiners are also known as the edge modes.

To find an upper bound on this entanglement measure we use the definition of $E_\tau^*$ and the inequality in (\ref{convexity}):  
\begin{eqnarray}\label{upper}
S^{\mF_{12}}(\omega_{12}\|E^*_\tau(\omega_{12}))\leq \log |G|
\end{eqnarray}
In section \ref{sec:5.2}, we demonstrate a generalization of the inequality (\ref{convexity}) that applies to QFT.

\subsection{Modular theory in the presence of charges}\label{sec:5.2}

\begin{figure}[h]
 \centering
 \begin{subfigure}{.4\textwidth}
   \centering
   \includegraphics[width=.8\linewidth]{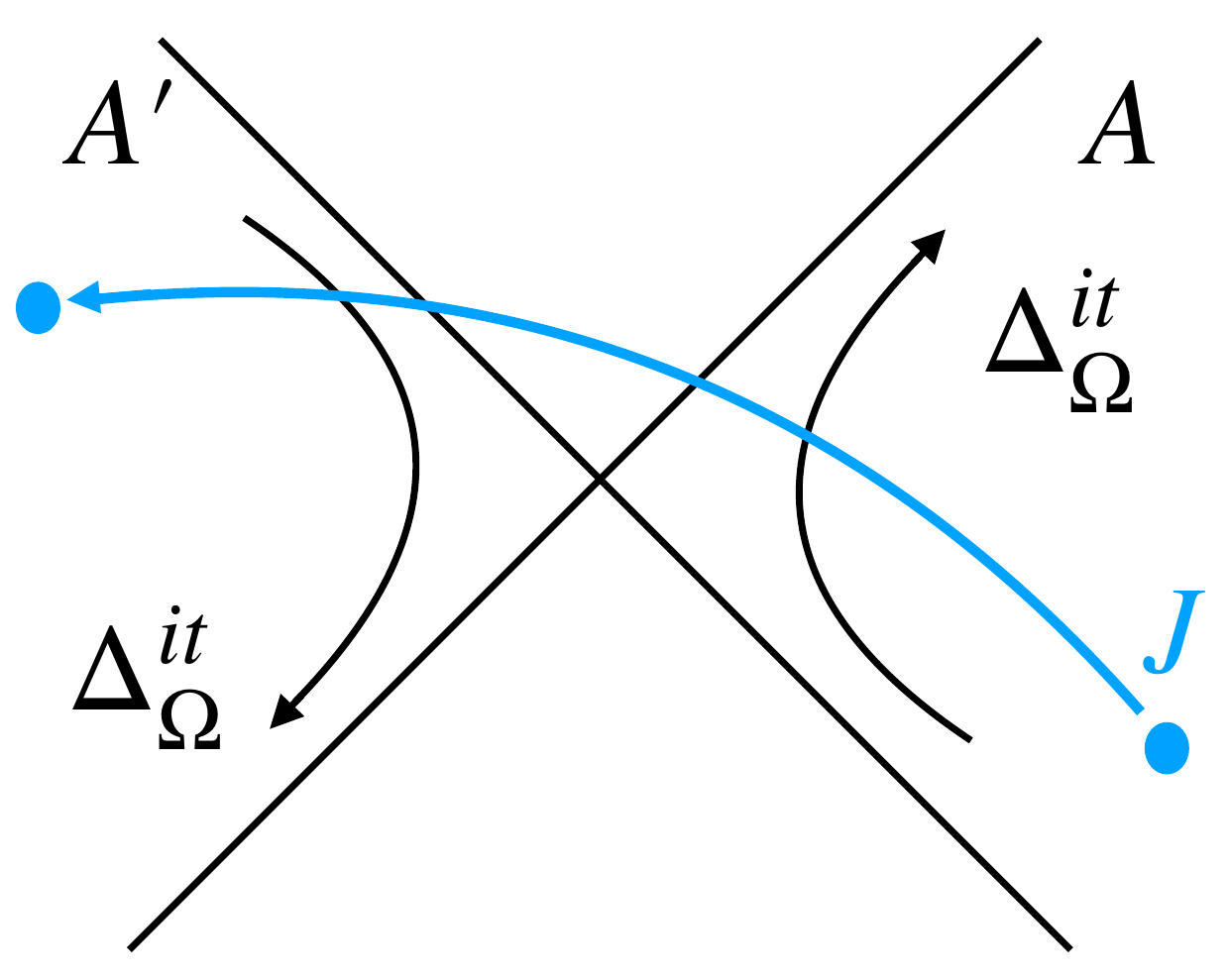}
   \caption{}
   \label{fig7:sub1}
 \end{subfigure}
 \begin{subfigure}{.4\textwidth}
   \centering
   \includegraphics[width=.8\linewidth]{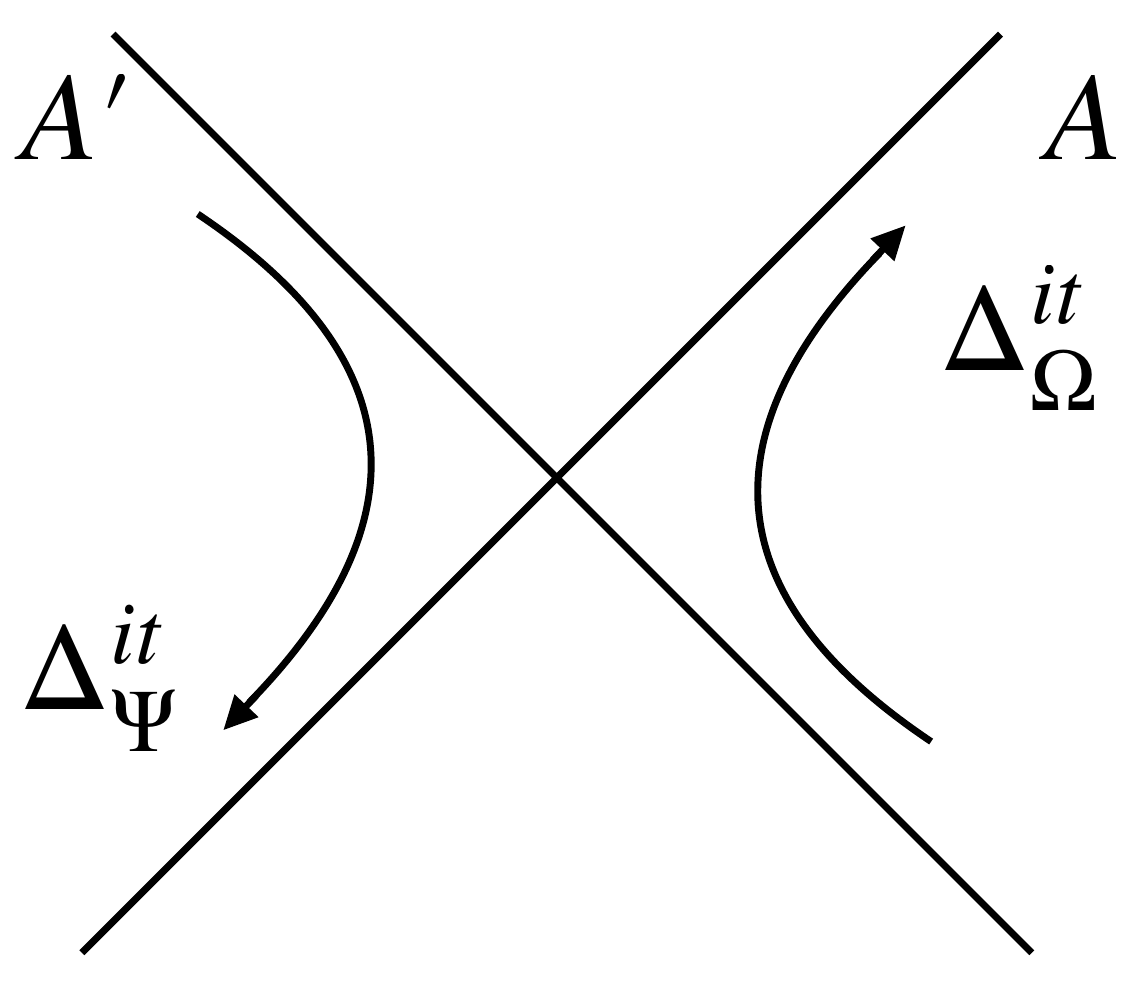}
   \caption{}
   \label{fig7:sub2}
 \end{subfigure}
 \caption{\small{If the region $A$ is the Rindler wedge and the state is a QFT in the vacuum, the modular flow is the boost that evolves operators geometrically according to the arrows in (a) \cite{witten2018aps}. The modular conjugation map $J_\Omega$ is the CRT (charge conjugation/reflection/time reversal) that sends operators from $A$ to $A'$ and vice versa. (b) The relative modular flow generated by $\Delta_{\Omega|\Psi}^{it}$ acts as the modular flow of $\Omega$ on the operators in $A$ and the modular flow of state $\Psi$ on the operators in $A'$.}}
 \label{fig7}
\end{figure}

Consider two global vectors of a QFT, $\ket{\Omega}$ and $\ket{\Psi}$ and a local algebra $\mF_A$. 
The relative Tomita operator is defined using the equation
\begin{eqnarray}\label{relativeTomita}
S^{\mF_A}_{\Psi|\Omega} b\ket{\Omega}=b^\dagger\ket{\Psi},\qquad \forall b\in \mF_A\ .
\end{eqnarray}
This operator is labelled by the choice of two vectors and an algebra. To simplify the notation, when it is clear from the context we suppress the algebra label.
The equation above defines the action of $S^{\mF_A}_{\Psi|\Omega}$ and its $\dagger$ everywhere in $\mH$ if the action of operators in $\mF_A$ and $\mF_{A'}$ on $\ket{\Omega}$ is dense in the Hilbert space: $\overline{\mF_A\ket{\Omega}}=\mH$ \cite{witten2018aps}. Such a vector $\ket{\Omega}$ is  called a Reeh-Schlieder vector (cyclic and separating). 
In a Reeh-Schlieder state, the action of local algebra $\mF_A$ on $\ket{\Omega}$ can approximate any excitation in the global Hilbert space, even those supported outside of $A$.\footnote{In finite quantum systems, the canonical purification of a density matrix $\rho$ is a Reeh-Schlieder vector if and only if all the eigenvectors of $\rho$ are non-zero. That is to say $\rho_A$ is entirely entangled with $A'$.} 
The vector $\ket{\Omega}$ is called Reeh-Schlieder if and only if it is cyclic with respect to both $\mF_A$ and $\mF_{A'}$.
The squared norm of the relative Tomita operator is called the relative modular operator $\Delta_{\Psi|\Omega}=S_{\Psi|\Omega}^\dagger S_{\Psi|\Omega}$ and we define the anti-linear operator $J_{\Psi|\Omega}=S_{\Psi|\Omega}\Delta_{\Psi|\Omega}^{-1/2}$. 
When both vectors are the same we call $S_{\Omega}\equiv S_{\Omega|\Omega}$ the Tomita operator and $\Delta_{\Omega}\equiv \Delta_{\Omega|\Omega}$ the modular operator. The anti-linear operator $J_\Omega=\Delta^{1/2}_\Omega S_{\Omega}$ is called the modular conjugation of $\ket{\Omega}$ and has the property that 
\begin{eqnarray}
 b_J\equiv J b J\in\mF_{A'}\qquad \forall b\in\mF_A,
\end{eqnarray}
where we have suppressed the $\Omega$ index of $J$. If $\ket{\Omega}$ is Reeh-Schlieder the modular conjugation is an anti-unitary $J=J^{-1}=J^\dagger$ \cite{bratteli1996operator}.
An important result of the modular theory is that the (relative) modular operator generates a flow called the (relative) modular flow that is an outer automorphism of the algebra $\mF_A$. This flow is independent of the second vector (for a proof see \cite{bratteli2012operator,lashkari2018modular}); see figure \ref{fig7}:
\begin{eqnarray}\label{relmodflow}
&&(\Delta_{\Omega|\Psi}^{\mF_A})^{it} b (\Delta_{\Omega|\Psi}^{\mF_A})^{-it}=(\Delta_{\Omega}^{\mF_A})^{it} b (\Delta_{\Omega}^{\mF_A})^{-it}\in \mF_A\qquad \forall b\in \mF_A \quad \text{and} \quad \forall t\in \mathbb{R}\nn\\ 
&&(\Delta_{\Omega|\Psi}^{\mF_A})^{it} b' (\Delta_{\Omega|\Psi}^{\mF_A})^{-it}=(\Delta_{\Psi}^{\mF_A})^{it} b' (\Delta_{\Psi}^{\mF_A})^{-it}\in \mF_A\qquad \forall b'\in \mF_{A'} \quad \text{and} \quad \forall t\in \mathbb{R}
\end{eqnarray}

The above relations imply that the operator $\Delta_{\Omega|\Psi}^{it}\Delta_\Omega^{-it}$ is in the commutes with all operators in $\mF_A$. 
This operator is called the cocycle. Similarly for the modular conjugation we have
\begin{eqnarray}
J_{\Psi|\Omega}^\dagger b J_{\Psi|\Omega} = J_\Omega b J_\Omega
\end{eqnarray}
which implies that $J_{\Psi|\Omega}J_\Psi$ commutes with all $\mF_A$. 
The correlation functions of the operators $b,c\in \mF_A$ in the state $\ket{\Omega}$ have the KMS property which can be interpreted as an analytic continuation of the modular flow to complex values of $t$: $\braket{\Omega|b\Delta_\Omega c|\Omega}=\braket{\Omega|c b|\Omega}$.\footnote{To show this we note that $\braket{\Omega|b\Delta c|\Omega}=\braket{\Omega|b S^\dagger S c|\Omega}=\braket{S c\Omega|S b^\dagger\Omega}=\braket{\Omega|c b|\Omega}$, where we have used the anti-linearity of $S$.} The set of operators $h\in \mF_A$ with the property that $\braket{\Omega| [h, b] |\Omega}=0$ for all $b\in\mF_A$ forms a subalgebra of $\mF_A$ that we call the centralizer of $\omega$ and denote it by $\mF_A^\omega$ \cite{pedersen1973radon,lashkari2019sewing}. The KMS property implies that 
\begin{eqnarray}
\braket{\Omega| b(\Delta-1)h |\Omega}=0\qquad \forall h\in \mF_A^\omega\ .
\end{eqnarray}
Since $b\ket{\Omega}$ is dense in the Hilbert space the vector $h\ket{\Omega}$ is an invariant state of the modular operator. 
The operators in the centralizer have the important property that $h$ and $\Delta$ commute \cite{pedersen1973radon}
\begin{eqnarray}\label{propcentralizer}
&&\Delta^z h \Delta^{-z}=h\qquad \forall z\in\mathbb{C}\ .
\end{eqnarray}
In fact, an operator $h\in \mF_A$ that is in the centralizer of $\Omega$ commutes with $\Delta_{\Omega|\Psi}$ for any $\Psi$.\footnote{To see this, we first rewrite $b$ as  $\lim_{\gamma\to \infty} b_\gamma$ in (\ref{bgamma}) that is entire meaning that $b_\gamma(z)$ defined in (\ref{mirror}) is in $\mF_1$ for all complex $z$. Then, from (\ref{relmodflow}) it follows that for all $h$ in the centralizer of $\Omega$ we have $\Delta_\Omega^z h \Delta_\Omega^{-z} =\Delta_{\Omega|\Psi}^z b \Delta_{\Omega|\Psi}^{-z}=h$.}
Since $h\in \mF_A^\omega$ are invariant under the modular flow, we sometimes refer to them as the modular zero modes. The modular zero mode satisfies the equation
\begin{eqnarray}\label{vectorcenter}
(h^\dagger-h_J)\ket{\Omega}=0\ .
\end{eqnarray}
Note that $h_J\in \mF_{A'}$ is also in the centralizer of $\Omega$.
If the algebra has a center $Z$, the center is inside the centralizer of all states. The operators in the center $z\in Z$ satisfy $z^\dagger=z_J$ \cite{araki2005extension}:
\begin{eqnarray}
z^\dagger b \ket{\Omega}=b z^\dagger\ket{\Omega}=S (z b^\dagger)\ket{\Omega}=J \Delta^{1/2}z b^\dagger\ket{\Omega} =z_J J\Delta^{1/2}b^\dagger\ket{\Omega}=z_J b \ket{\Omega}\ .
\end{eqnarray}

The relative Tomita operator for an excited state $h\ket{\Omega}$ and $h$ an invertible element of the centeralizer is
 \begin{eqnarray} 
 &&S_{\Omega|h\Omega}=\|h\ket{\Omega}\|S_\Omega (h_J)^{-1}\nn\\
 &&\Delta_{\Omega|h\Omega}=\|h\ket{\Omega}\|^2 \Delta_\Omega |h_J|^{-2}
 \end{eqnarray}
 where we have used (\ref{vectorcenter}).
The relative entropy of two vectors with respect to an algebra $\mF_A$ is given by \cite{Araki:1976zv}
\begin{eqnarray}\label{Arakiform}
S^{\mF_A}(\Psi\|\Omega)=-\braket{\Psi|\log\Delta^{\mF_A}_{\Omega|\Psi}\Psi}\ .
\end{eqnarray}
When $\ket{\Omega}$ and $\ket{\Psi}$ are the canonical purifications of density matrices $\sigma$ and $\rho$ in (\ref{canonicalform}) the formula above matches the definition:
\begin{eqnarray}
S(\rho\|\sigma)=\text{tr}(\rho\log\rho)-\text{tr}(\rho\log\sigma)\ .
\end{eqnarray}
The elements of the centralizer are the operators that commute with the density matrix. The local state associated with the excited state $h\ket{\Omega}$ with $h$ in the centralizer is $\rho_h=h \rho h^\dagger/\text{tr}(\rho |h|^2)$ that commutes with $\rho$ the local state of $\ket{\Omega}$. The relative entropy of these states with respect to the vacuum defined by (\ref{Arakiform}) is
\begin{eqnarray}\label{relentropycentralizer}
S(h\Omega\|\Omega)=-2\log\|h\ket{\Omega}\|+2\braket{h\Omega|\log|h_J||h \Omega}
\end{eqnarray}
where $\ket{h \Omega}$ is the normalized state $h\ket{\Omega}$. 
If $v$ is an isometry in the centralizer of $\Omega$ then $v^\dagger\ket{\Omega}$ has the same local state as $\ket{\Omega}$:
\begin{eqnarray}
\braket{\Omega|v b v^\dagger|\Omega}=\braket{\Omega|v^\dagger v b|\Omega}=\braket{\Omega|b|\Omega}\ .
\end{eqnarray}
That is why the equation (\ref{relentropycentralizer}) implies $S(v^\dagger\Omega\|\Omega)=0$ for $v$ in the centralizer.
Since $\rho$ and $\rho_h$ are simultaneously block diagonalizable  their relative entropy can be understood as a classical relative entropy. For instance, take $\rho=\sum_k q_k\ket{k}\bra{k}$ and $h=\sum_k \sqrt{\frac{p_k}{q_k}}\ket{k}\bra{k}$ with $p_k$ a probability distribution that is in the centralizer of $\rho$. The state $\rho_h=\sum_k p_k\ket{k}\bra{k}$ is simultaneously diagonalized with $\rho$. The relative entropy above is
\begin{eqnarray}
S(\sum_k p_k\ket{k}\bra{k}\|\sum_k q_k\ket{k}\bra{k})=\sum_k p_k (\log p_k-\log q_k)=H(p\|q)
\end{eqnarray}
which is a special case of (\ref{relentropyadd}). 
More generally, for an operator $h$ that in the centralizer of $\ket{\Omega}$ we have
\begin{eqnarray}\label{centerprop}
 &&S_{\Psi|h\Omega}=\|h\ket{\Omega}\| 
  S_{\Psi|\Omega}(h_{J_\Omega})^{-1}\nn\\
 &&\Delta_{\Psi|\Omega}=\|h\ket{\Omega}\|^2 \Delta_{\Psi|\Omega} |h_{J_\Omega}|^{-2}
 \end{eqnarray}
 where we have used the fact that $[\Delta_{\Psi|\Omega},h_{J_\Omega}]=0$ because $h_{J_\Omega}\in \mF_{A'}$ and in the centralizer of $\Omega$.
 Then, the relative entropy is
 \begin{eqnarray}
 &&S(h\Omega\|\Psi)=-2\log\|h\ket{\Omega}\| +2\braket{h\Omega|\log|h_J||h\Omega}-\braket{h\Omega| \log\Delta_{\Psi|\Omega}|h\Omega}\ .
 \end{eqnarray}
  This is a QFT generalization of the equation (\ref{relentropyadd}). To see this, plug in the equation above the block diagonal density matrices $\rho=\oplus_k q_k \rho_k$, $\psi=\oplus_k p_k \sigma_k$ and the operator $h=\oplus_k \sqrt{\frac{p_k}{q_k}}\mathbb{I}_k$ that is the centralizer of both states: 
  \begin{eqnarray}
S^{\oplus_k \mF_k}(\rho\|\sigma)=H(p\|q)+\sum_k p_k S(\rho_k\|\sigma_k)\ .
\end{eqnarray}

In the presence of an internal symmetry $U_gbU_g^\dagger\in\mF_A$ for all $b\in\mF_A$. From (\ref{relativeTomita}) we can solve for the modular operator of $U_g\ket{\Psi}$:
\begin{eqnarray}\label{commuteModOp}
&&S^A_{U_g\Psi}=U_g S^A_{\Psi}U_g^\dagger\nn\\
&&\Delta^A_{U_g\Psi}=U_g\Delta^A_\Psi U_g^\dagger\ .
\end{eqnarray}
If $\ket{\Omega}$ is the invariant vacuum, i.e. $U_g\ket{\Omega}=\ket{\Omega}$, the modular operator and $U_g$ commute: $\Delta_\Omega U_g=U_g\Delta_\Omega$.
As a result, the modular flow $\Delta_\Omega$ of charge-neutral operators remains charge-neutral if $\ket{\Omega}$ is an invariant vector, and the charge of an operator $V_{r,i}$ does not change under the modular flow by $\Delta_\Omega^{it}$. Now, consider the twist unitary $\tau_g$.
On a lattice, the twist operator is in the center of the local charge-neutral algebra.
In QFT the twist operator is not in the center of the local algebra, but we still have
\begin{eqnarray}
&&S^{\mA_1\otimes \mA_2}_{\Psi|\tau_g\Omega} =S^{\mA_1\otimes\mA_2}_{\Psi|\Omega}\tau_g^\dagger\nn\\
&&S^{\mA_1\otimes \mA_2}(\tau_g\Omega\|\Psi)=S^{\mA_1\otimes \mA_2}(\Omega\|\Psi)\ .
\end{eqnarray}
This is expected because $\tau_g\ket{\Omega}$ has the same local state as $\ket{\Omega}$ with respect to the algebra $\mA_1\otimes \mA_2$. 

The relative Tomita operator of the charged states $\ket{r,i}=\sqrt{d_r}V_{r,i}^\dagger\ket{\Omega}$ is
\begin{eqnarray}
&&S_{(r'i')|(ri)}b\ket{r,i,A}=b^\dagger \ket{r',i',A},\forall b\in \mF_A
\end{eqnarray}
which can be solved by setting $S_{(r'i')|(ri)}= \sqrt{\frac{d_{r'}}{d_{r}}}V_{r,i}^\dagger S_\Omega V_{r',i'}$ in the equation above. Below, we suppress the algebra label in the relative modular operator and relative entropies if the algebra is $\mF_A$. 
Note that $S_\Omega$ is the Tomita operator for all charged operators, and we have used $V_{ri}^\dagger V_{rj}=\delta_{ij}$. The relative Tomita operator kills the vectors $b\ket{s,j,A}$ for $s\neq r$, and on its domain it satisfies
\begin{eqnarray}
&&S_{(r'i')|(ri)}= \sqrt{\frac{d_{r'}}{d_{r}}}V_{r,i}^\dagger S_\Omega V_{r',i'}\nn\\
&&\Delta_{(r'i')|(ri)}= \frac{d_{r'}}{d_{r}} V_{r',i'}^\dagger S_\Omega^\dagger V_{r,i} V_{r,i}^\dagger S_\Omega V_{r',i'}\ .
\end{eqnarray}
In particular, we find that $\sum_i\Delta_{(r'i')|(ri)}=\frac{1}{d_{r}}\Delta_{(r'i')|\Omega}$.
For an Abelian symmetry, the intertwiner $V_q$ is a unitary operator and 
\begin{eqnarray}
\Delta_{q|q'}=V_q^\dagger \Delta_\Omega V_q\ .
\end{eqnarray}
Therefore, the relative entropy states of sectors of charge $q$ and $q'$ is
\begin{eqnarray}
S^{\mF_A}(q'\|q)=-\braket{q'|V_q^\dagger\log\Delta_\Omega V_q|q'}=-\braket{q'-q|\log\Delta_\Omega|q'-q}\ .
\end{eqnarray}
For bi-local intertwiners $\mcI_{12,r}$ and the algebra $\mF_{12}$ we have
\begin{eqnarray}
&&S^{\mF_{12}}_{\Omega|\mcI_r}=\mcI_r S^{\mF_{12}}_\Omega\nn\\
&&S^{\mF_{12}}_{\mcI_r|\Omega}=S^{\mF_{12}}_\Omega \mcI_r^\dagger\nn\\
&&S^{\mF_{12}}(\mcI_r\|\Omega)=-\braket{\mcI_r|\log\Delta_\Omega|\mcI_r}\nn\\
&&S^{\mF_{12}}(\Omega\|\mcI_r)=-\braket{\mcI^\dagger_r|\log\Delta_\Omega|\mcI^\dagger_r}\ .
\end{eqnarray}
where for $\mF_1$
\begin{eqnarray}
&&S^{\mF_1}_{\Omega|\mcI_r}=\frac{1}{d_r}\sum_j (V_{r,j}^{(1)})^\dagger S_\Omega^{\mF_1}(V_{r,j}^{(2)})^\dagger,\nn\\
&&S^{\mF_1}_{\mcI_r|\Omega}=\sum_j (V_{r,j}^{(2)}) S_\Omega^{\mF_1}(V_{r,j}^{(1)})\ .
\end{eqnarray}

The relative Tomita equation defines the relative modular operator unambiguously if the vector $\ket{\Omega}$ is Reeh-Schlieder. The Poincare-invariant vacuum of QFT is a Reeh-Schlieder vector for local algebras $\mF_A$. In a Reeh-Schlieder vector the excitations inside the region $A$ can approximate an arbitrary excitations outside. We are interested in studying the relative modular operator with respect to the local charge-neutral subalgebras $\mA_A$, and below we show that the vacuum vector is Reeh-Schlieder with respect to $\mA_A$.
That is to say in QFT an arbitrary uncharged operator in $\mA$ can be approximated using local uncharged operators $\mA_A$: $\overline{\mA_A\ket{\Omega}}=\overline{\mA\ket{\Omega}}$. 

First, let us take a look at the Reeh-Schlieder property for the full algebra $\mF$ of QFT. 
In a Reeh-Schlieder state an arbitrary excitation in $\mF_1\otimes \mF_{1'}$ can be approximated using operators in $\mF_1$.
We want to find $b_m\in \mF_1$ such that for some $b'\in \mF_{1'}$ we have $b'\ket{\Omega}\simeq b_m\ket{\Omega}$. We call such an operator $b_m$ the {\it mirror operator} of $b'$. 
To construct the mirror operator, we use the following strategy
\begin{eqnarray}
b'\ket{\Omega}=S'_{\Omega} (b')^\dagger\ket{\Omega}=(\Delta')^{-1/2}J(b')^\dagger \ket{\Omega}= \Delta^{1/2}(b')^\dagger_J\Delta^{-1/2}\ket{\Omega}
\end{eqnarray}
with $(b')^\dagger_J\equiv J (b')^\dagger J\in \mF_1$ where we have suppressed the $\Omega$ index of $\Delta_\Omega$ and $J_\Omega$. For a Reeh-Schlieder vector in finite quantum systems, it is straightforward to check that the operator $(\omega^{1/2} b^T\omega^{-1/2}\otimes 1)$ is the mirror of $(1\otimes b)$ where $T$ is the transpose in the basis picked by the density matrix $\omega$:
\begin{eqnarray}
\sum_k \sqrt{p_k}(\omega^{1/2} b^T\omega^{-1/2}\otimes 1)\ket{kk}=\sum_{kl}\sqrt{p_l}b_{kl}\ket{lk}=\sum_l\sqrt{p_l}(1\otimes b)\ket{ll}\ .
\end{eqnarray}
where we have used the canonical purification of $\omega$ in (\ref{canonicalform}).
Note that in the example above, the modular conjugation operator $J$ is the anti-linear swap operator in the Schmidt basis of the state: $J c \ket{kl}=c^*\ket{lk}$ where $c$ is a complex number.
In a Reeh-Schlieder state since all $p_k>0$, $\omega^{-1/2}$ is well-defined. Furthermore, the operator $\Delta^{1/2}(b^T\otimes 1)\Delta^{-1/2}=\omega^{1/2}b^T\omega^{-1/2}\otimes 1\in \mF_1\otimes 1$.  
In a QFT, for a general $b\in\mF_1$, the modular flow $b(t)\equiv \Delta^{it}b\Delta^{-it}$ is inside the algebra $\mF_1$ for all $t\in\mathbb{R}$, but the operator $\Delta^{1/2}b\Delta^{-1/2}$ need not be in $\mF_1$. 
Luckily, as we demonstrate below, in QFT there are always operators in $\mF_1$ that approximate $\Delta^{1/2}b \Delta^{-1/2}$ arbitrarily well.  

\begin{figure}[h]
 \centering
 \begin{subfigure}{.4\textwidth}
   \centering
   \includegraphics[width=.8\linewidth]{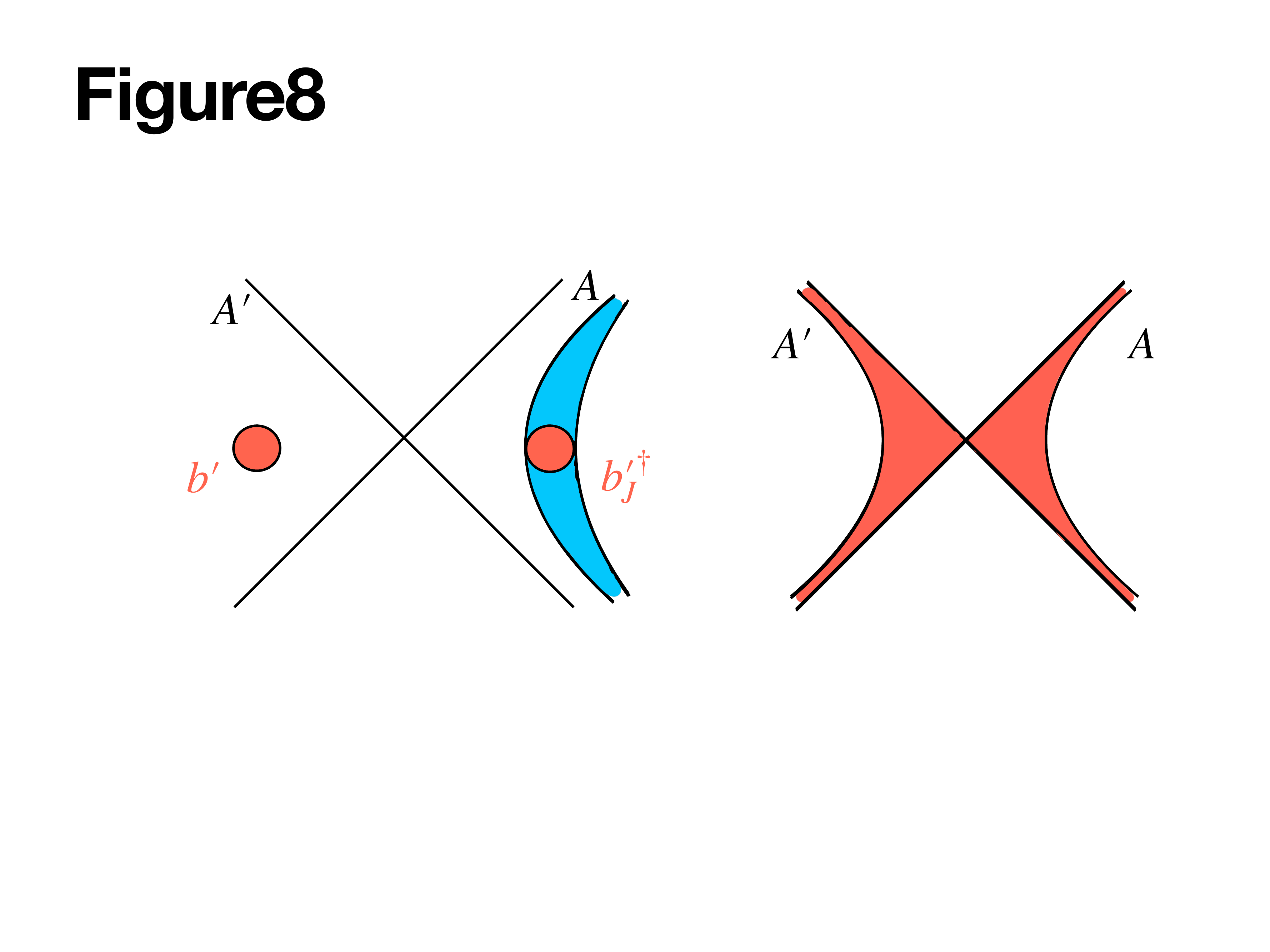}
   \caption{}
   \label{fig8:sub1}
 \end{subfigure}
 \begin{subfigure}{.4\textwidth}
   \centering
   \includegraphics[width=.8\linewidth]{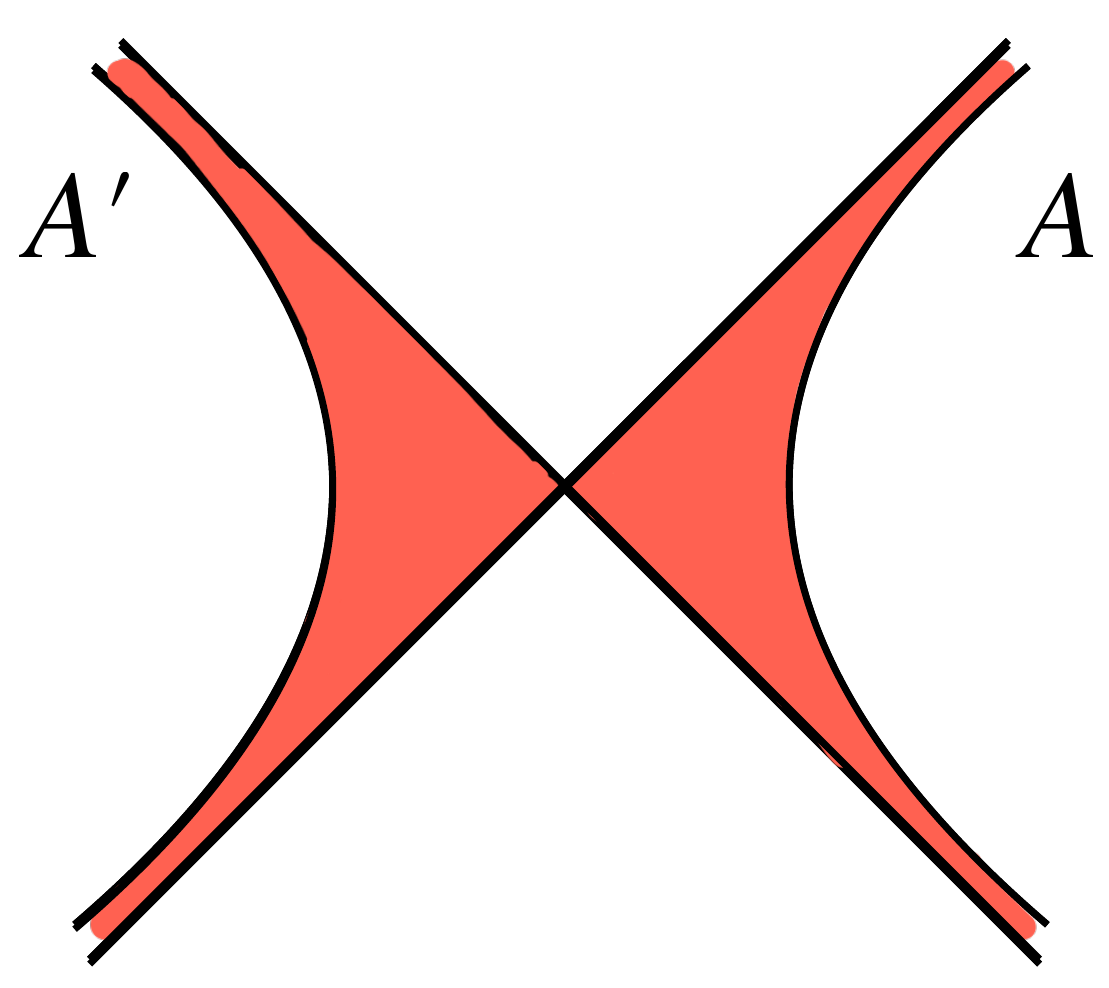}
   \caption{}
   \label{fig8:sub2}
 \end{subfigure}
 \caption{\small{Consider the case where region $A$ is the Rindler wedge and we have the vacuum of QFT (a) If $b'$ is localized in the small red circle inside $A'$ the operator $(b'_J)^\dagger$ is also localized in the red circle in $A$, however the the mirror operator in (\ref{mirror}) requires boosting that spreads its support in the blue region (b) The operators in $A'$ that are approximately-invariant under modular flow (boost) are localized in a small proper distance from the entangling surface. Their mirror operators are also localized near the entangling surface in $A$.}}
 \label{fig8}
\end{figure}

Consider the operator
\begin{eqnarray}\label{bgamma}
b_\gamma=\sqrt{\frac{\gamma}{\pi}}\int_{-\infty}^\infty dt \:e^{-\gamma t^2} \Delta^{it}b\Delta^{-it}\in \mF_1
\end{eqnarray}
In the limit $\gamma\to\infty$ this operator approximates $b$,\footnote{Note that in the limit $\gamma\to 0$ the operator $b_0$ is the modular zero mode, and for finite values of $\gamma$ this operator sends off-diagonal elements  $\ket{k}\bra{k'}\to e^{-\frac{(\log p_k-\log p_{k'})^2}{4\gamma}}\ket{k}\bra{k'}$. It suppresses the off-diagonal terms exponentially with parameter $\frac{1}{\gamma}$. The modular zero mode has the property that its modular flow is trivial: $(b_0)_\gamma=b_0$ for all $\gamma$.} and for any $\gamma$ the modular flow of this operator can be analytically continued to the whole complex plane \cite{pedersen1973radon}
\begin{eqnarray}\label{mirror}
b_{\gamma}(z)\equiv \Delta^{z} b_\gamma\Delta^{-z}=\sqrt{\frac{\gamma}{\pi}}\int_{-\infty}^\infty dt \:e^{-\gamma (t+i z)^2} \Delta^{it}b\Delta^{-it}\in \mF_1\ .
\end{eqnarray}
Therefore, we find that mirror operator of $b'$ in the algebra $\mF_1$ that satisfies
\begin{eqnarray}\label{mirroreq}
&& b'\ket{\Omega}\simeq b_m\ket{\Omega}\nn\\
 &&b_m=\lim_{\gamma\to \infty}((b')^\dagger_J)_\gamma(1/2)\ .
\end{eqnarray}
If the operator $b'$ is an isometry, the equation 
\begin{eqnarray}
\braket{\Omega|b'b_m^\dagger\Omega}=1
\end{eqnarray}
implies that the probability for the spontaneous creation of the excitation $b' b_m^\dagger\ket{\Omega}$ is almost one. In general, if $b'$ is localized in a small region of $A_{1'}$ its mirror is highly delocalized in $A_{1}$. 
If $[b',\Delta]=0$ from the mirror equation (\ref{mirroreq}) we find that $b_m=(b')^\dagger_J$; see figure \ref{fig8}.

Consider the symmetry group $G$ acting on the global algebra $\mF$.
If $a'\in\mA_{1'}$ is a charge-neutral operator from (\ref{mirror}) it is evident that the mirror operator $a_m$ is also charge-neutral, and is therefore in $\mA_1$. This implies that we can generate $\mA_{1'}\ket{\Omega}$ using $\mA_1\ket{\Omega}$.
The only other operators in $\mA$ are the bi-local intertwiners between $A$ and $A'$: $I_{r}=\sum_i V_{r,i}^\dagger V'_{r,i}$. Denote the mirror of $V'_{r,i}$ by $(V_{r,i})_m$. It has the same charge as $V'_{r,i}$. Therefore, the operator $\sum_i V_{r,i}^\dagger (V_{r,i})_m$ is charge-neutral with respect to the local algebra and therefore belongs to $\mA_A$. Moreover, the mirror of all operators in $\mA_{1'}$ also belong to $\mA_1$, therefore 
in QFT $\overline{\mA_1\ket{\Omega}}=\overline{\mA\ket{\Omega}}$. 

Since $\overline{A_1\ket{\Omega}}=\overline{\mA_{12}\ket{\Omega}}$, it follows that for non-overlapping and non-touching $A_1$ and $A_2$ we have $\overline{\mA_1\otimes \mA_2\ket{\Omega}}=\overline{\mA_{12}\ket{\Omega}}=\overline{\mA\ket{\Omega}}$. All intertwiners between region $A_1$ and the complement $A'_1$ can be prepared locally by acting with $\mA_1$ which includes the intertwiners between regions $A_1$ and $A_2$. As a result, the algebra $\overline{(\mA_1\otimes \mA_2)\ket{\Omega}}=\overline{\mA\ket{\Omega}}$, and the Tomita operator for the algebra $\mA_1\otimes \mA_2$ is densely defined.

\subsection{Cocycle and intertwiners}\label{sec:5.3}

In this section, we show that the intertwiner $V_{r,i}$ can be understood as an analytic continuation of the unitary cocycle. Consider two vectors $\ket{\Omega}$ and $\ket{\Psi}$ in different superselection sectors of a QFT and the isometry defined by 
\begin{eqnarray}\label{isom}
T(a\ket{\Psi})=a\Delta^{1/2}_{\Psi|\Omega}\ket{\Omega} 
\end{eqnarray}
that maps vectors from the $\ket{\Omega}$ to the $\ket{\Psi}$ sector. This is an intertwiner that takes us from one charged sector to another and commutes with the action of $ \mF_A$ \cite{witten2018aps}.
When the superselection sectors are due to symmetries, the intertwiner need not be localized in $A$. 
We say the intertwiner is localized in $A$ if $T b'\ket{\Omega}=b'T\ket{\Omega}$ for all $b'$ charged operators in $A'$. We would like to understand when the intertwiner $T$ is localized in $A$. 
In the last subsection, we saw that the cocycle operator 
\begin{eqnarray}
u_{\Psi|\Omega}(t)=\Delta_{\Psi|\Omega}^{it}\Delta_\Omega^{-it}
\end{eqnarray}
belongs to the algebra. In fact, if both $\Omega$ and $\Psi$ are Reeh-Schlieder it is a unitary operator.
For real values of $t$ the cocycle is an operator in $A$ and commutes with $b'$ \cite{connes1973classification,lashkari2019sewing}.
The isometry in (\ref{isom}) can be created by an analytic continuation of the cocycle to imaginary values $\Im(t)=-i/2$:
\begin{eqnarray}
T b'\ket{\Psi}=b'\Delta_{\Psi|\Omega}^{1/2}\Delta_\Omega^{-1/2}\ket{\Omega}=b' u_{\Psi|\Omega}(-i/2)\ket{\Omega}\ .
\end{eqnarray}
The isometry in (\ref{isom}) commutes with all $b'\in F_{A'}$ if the analytic continuation of the cocycle $u_{\Psi|\Omega}(-i/2)$ exists and belongs to $\mF_A$.

On a lattice the cocycle is $u_{\Psi|\Omega}(t)=\psi^{it}\omega^{-it}\otimes 1'$ with $\psi$ and $\omega$ the reduced density matrices on $A$ of $\ket{\Psi}$ and $\ket{\Omega}$, respectively. The analytic continuation of the cocycle to $\Im(t)=-i/2$ corresponds to $\psi^{1/2}\omega^{-1/2}\otimes 1$ which is well-defined if all the density matrix $\omega$ has no zero eigenvalues. In fact, it suffices to assume that every zero eigenvalue of $\omega$ is also a zero eigenvalue of $\psi$, because if $\omega\ket{\xi}=\psi\ket{\xi}=0$ we can define 
$\psi^{1/2}\omega^{-1/2}\ket{\xi}=0$.
In other words, the cocycle has an analytic continuation if there exists a $\lambda>0$ such that $\omega-\lambda \psi$ is a non-negative operator. This is the necessary condition for the relative entropy $S(\psi\|\omega)$ to be finite. Similarly, in modular theory, the cocycle $u_{\Psi|\Omega}(t)$ can be analytically continued to the $0\geq \Im(t)\geq -1/2$ if $\omega-\lambda \psi\geq 0$. That is to say there exists a $\lambda>0$ such that for all $b\in \mF_A$:
\begin{eqnarray}
\omega(b^\dagger b)-\lambda \psi(b^\dagger b)\geq 0\ .
\end{eqnarray}

If $\ket{\Psi}$ and $\ket{\Omega}$ are vectors corresponding to states $\psi$ and $\omega$ with $\omega\geq \lambda \psi$ for some positive $\lambda$,
the cocycle $u_{\psi|\omega}$ has an analytic continuation to the strip $0\geq \Im(t)\geq -1/2$ that remains inside the algebra \cite{araki1976relative}. This implies that there is a map $u(t)$ analytic in the strip and strongly continuous in $t$ with the property that  
\begin{eqnarray}
&&u(t)=u_{\psi|\omega}(t)\nn\\
&&u(t)b'\ket{\Omega}=b'\Delta_{\Psi|\Omega}^{it}\ket{\Omega}\ .
\end{eqnarray}
In particular at $u(-i/2)$ we have the map
\begin{eqnarray}
u(-i/2)b'\ket{\Psi}=b'u(-i/2)\ket{\Psi},
\end{eqnarray}
which is the local intertwiners $V_{r,i}$ we discussed in the case of charges in section \ref{sec:4}. 
In our examples of QFT with charges we have $E^*_\tau(\omega_{12})>\frac{1}{|G|}\omega_{12}$, therefore the cocycle $u_{\omega_{12}|E^*_\tau(\omega_{12})}$ takes us from the sector $\ket{\Omega}$ to the sector corresponding to $E_{\tau}^*(\omega)$.

\subsection{Enlarging the QFT algebra}\label{sec:5.4}

In section \ref{sec:4}, we saw that one main difference between QFT and systems on a lattice is that in QFT the twist operator $\tau_g=e^{i g Q_A}$ is not part of the local algebra, and as a result the local invariant algebra has no center.
It is natural to ask whether one can enlarge the QFT algebra by including $\tau_g$ to make QFT more similar to the lattice models.
The local algebra of charged operators in QFT has charge neutral operators $\mI_r\otimes a \in \mA_A$ and charged operators $V^{(A)}_{r,i}$ supported on $A$ that belong to the dual group $\hat{G}$: $\mF_A=\mA_A\rtimes_\rho \hat{G}$. If we further enlarge the QFT algebra by adding $\tau_g$ that belongs to $G$ to it, we obtain $\mF_A\rtimes G$. If the group $G$ is Abelian this is $(\mA_A\rtimes \hat{G})\rtimes G=\mA_A\otimes B(L^2(\hat{G}))$. Physically, this corresponds to adding a qudit of dimension $|G|$ to the local algebra of QFT exactly as we do on a lattice \cite{pedersen1973radon}. There will be an analogous degree of freedom on the complementary region $A'$ and the global Hilbert space factors as
$\mH=\oplus_q(\mK^A_q\otimes \mK^{A'}_q)\otimes \mH_q$. 
The enlarged local algebra is the tensor product of the algebra of charge neutral operators with a qudit of dimension $|G|$ that carries the charge: $\mA_1\otimes GL(|G|,\mathbb{C})$ where $GL(|G|,\mathbb{C})$ is the algebra of a qudit. In this enlarged algebra, the charge neutral operators have a non-trivial center: $\oplus_q \lambda_q \ket{q}\bra{q}\otimes 1$ similar to systems on a lattice. 
When the algebra is non-commutative, it is still convenient to consider the algebra represented on the Hilbert space $\oplus_r (\mK_r^{A}\otimes \mK_r^{A'})\otimes \mH_r$. 

It is desirable to construct a conditional expectation that maps the enlarged algebra $\mF_A\rtimes G$ back down to the QFT algebra $\mF_A$. In the Abelian case, this is simply an average over the dual group
\begin{eqnarray}
\tilde{E}(\ket{q'}\bra{q''}\otimes a)=\frac{1}{|\hat{G}|}\sum_q \hat{U}_q^\dagger (\ket{q'}\bra{q''}\otimes a)\hat{U}_q\ .
\end{eqnarray}
The QFT local algebra $\mF_A$ is generated by $\ket{0}\bra{0}\otimes a$ and $\hat{U}_q$ which transform under this map to
\begin{eqnarray}\label{transf}
&&\tilde{E}(\ket{0}\bra{0}\otimes a)=\frac{1}{|\hat{G}|}\sum_q\rho_q(a)\nn\\
&&\tilde{E}(\hat{U}_{q})=\hat{U}_q
\end{eqnarray}
where in the second line we have used the fact that the group is Abelian.
Under $\tilde{E}$ any new non-identity elements of the twist group $\tau_g = \sum_q e^{igq} \ket{q,A}\bra{q,A}$ are washed out
\begin{eqnarray}
\tilde{E}(\tau_g)= \delta_{g 0} \mathbb{I}\ .
\end{eqnarray}
In the example of the free boson in section \ref{sec:3.5}, the dual group is $\mathbb{Z}$ that is not compact but we can still write
\begin{eqnarray}
\rho_\alpha(J(u))=J(u)+\oint\frac{dz}{2\pi i} u(z)\alpha(z)
\end{eqnarray}
and the sum over charges in the range $(-q,q)$ vanishes
\begin{equation}
\frac{1}{2q+1}  \sum_{k=-q}^q \rho_{k\alpha}(J(u))=J(u)\ .
\end{equation}
as one expects from a conditional expectation.

Our enlarged algebra has a representation in a Hilbert space that factors the charge modes $\oplus_q (\mK_q^A\otimes \mK^{A'}_q)\otimes \mH_q$. 
We would like to find the vectors in this Hilbert space that correspond to the states of the QFT algebra. We can extend our QFT states using the conditional expectation $\tilde{E}$ so that the relative entropy of states evaluated in the enlarged algebra remains the same as that of the QFT algebra.
For instance, the purification of the state $E^*(\psi)$ in (\ref{purification}) in this enlarged Hilbert space is 
\begin{eqnarray}
&&\ket{\Theta_\psi}=\sum_r \sqrt{\zeta_r}\ket{E_r}\otimes \ket{\Psi_r}\nn\\
&&\ket{E_r}=\frac{1}{\sqrt{d_r}}\sum_i \ket{r,i}\otimes\ket{r^*,i}
\end{eqnarray}
where $r^*$ is the dual representation of $r$.
The expectation values in this vector are
\begin{eqnarray}
 &&\sum_r\braket{\Theta_\psi|(1_r\otimes a)|\Theta_\psi}=  \sum_r \zeta_r\rho_r(a)\nn\\
 &&\braket{\Theta_\psi|V_{r,i}|\Theta_\psi}=0
\end{eqnarray}
and
\begin{eqnarray}
\braket{\Theta_\psi|U^r_g\otimes 1|\Theta_\psi}=\frac{1}{d_r}\text{tr}(U^r_g)=\delta(g)
\end{eqnarray}
as expected from an invariant state of $\tilde{E}$.
This vector is also Reeh-Schlieder with respect to the QFT algebra because the action of $V_{r,i}$ and $\mI_r\otimes a$ take us everywhere in the Hilbert space. 

For simplicity, we assume that the symmerty group is Abelian for the remainder of this section.
The relative Tomita equation for the vectors $\ket{\Theta_\omega}=\ket{00}\otimes \ket{\Omega}$ and $\ket{\Theta_\psi}=\ket{00}\otimes \ket{\Psi}$ is
 \begin{eqnarray}
 &&S^\mF_{\Theta_\psi|\Theta_\omega}(\ket{q}\bra{q'}\otimes a)\ket{\Theta_\omega}=(\ket{q'}\bra{q}\otimes a^\dagger)\ket{\Theta_\psi}\nn\\
 &&S^\mA_{\Theta_\psi|\Theta_\omega}(\mathbb{I}\otimes a)\ket{\Theta_\omega}=(\mathbb{I}\otimes a^\dagger)\ket{\Theta_\psi}\ .
 \end{eqnarray}
 The domain of $S^\mF$ is $\ket{q0}\otimes\mH_0$, however, this operator is zero except for the subspace $\ket{00}\otimes \mH_0$ that is the domain of $S^\mA$. On the common domain the two relative modular operators agree. Since the zero vector is not in the domain of $(S^\mF)^\dagger$ the relative modular operators $\Delta^\mF$ and $\Delta^\mA$ are the same map from $\ket{00}\otimes\mH_0\to \ket{00}\otimes \mH_0$. 
  In fact, the purification of any state that is invariant under the conditional expectation $E$ has this property. 
 The distinguishability of invariant states does not change as the restriction map $E$. Consider two invariant states:
\begin{eqnarray}
&&\omega=\sum_qp_q\ket{q}\bra{q}\otimes \omega_q\nn\\
&&\psi=\sum_qp'_q\ket{q}\bra{q}\otimes \psi_q
\end{eqnarray}
and their corresponding purifications
\begin{eqnarray}
&&\ket{\Theta_\omega}=\sum_q \sqrt{p_q}\ket{q,-q}\otimes \ket{\Omega_q}\nn\\
&&\ket{\Theta_\psi}=\sum_q \sqrt{p'_q}\ket{q,-q}\otimes \ket{\Psi_q}
\end{eqnarray}
Their relative modular operator is
\begin{eqnarray}
\Delta^\mF_{\Theta_\psi|\Theta_\omega}=\Delta^\mA_{\Theta_\psi|\Theta_\omega}=\sum_{l,m}\frac{p'_l}{p_m}\ket{l,-m}\bra{l,-m}\otimes\Delta_{\Psi_l|\Omega_m}
\end{eqnarray}
and their relative entropy is
\begin{eqnarray}
S(\Theta_\omega\|\Theta_\psi)=-\braket{\Theta_\omega|\log\Delta_{\Theta_\psi|\Theta_\omega}\Theta_\omega}=H(p\|p')+\sum_l p_l S(\Psi_l\|\Omega_l)
\end{eqnarray}
as expected from the equation (\ref{convexity}).

The algebra of QFT does not admit a tensor factorization when the regions $A_1$ and $A_2$ touch, however, as we saw in the presence of a symmetry the extended algebra factors the charged excitations. The local algebra of any quantum field has a symmetry group $\mathbb{R}$ associated with the modular flow. 
The modular flow is an outer automorphism similar to the twist group. Similar to the case of twist that was not part of the algebra due to the infinities near the entangling surface, the modular Hamiltonian, i.e. $\log\Delta_{\Omega}$ restricted to $A$ is not part of the algebra because of its discontinuous action at the entangling surface. For instance, in the vacuum QFT and for the Rindler region $|x^1|>|t|$ the modular Hamiltonian is the boost operator $\int_{-\infty}^\infty du\: u T_{uu}$ where $T_{uu}$ is the null-null component of the stress tensor. The half-sided modular Hamiltonian $\int_0^\infty du \: u T_{uu}$ is ill-defined because of its singular behavior at $u=0$.
If we enlarge the local algebra of QFT by the modular group by adding the half-sided modular Hamiltonian to the algebra, every mode that is charged under modular flow factors. 
The Hilbert space splits into sectors $\mH_q$ with projections 
$P_q$ that project to the subspace with modular frequency $q$. 

The modular group is $\mathbb{R}$ so its dual group is also $\mathbb{R}$ which is non-compact. 
In the case of vacuum QFT in Rindler space, the centralizer is trivial since there are no local operators that are invariant under boost. This implies that every mode is charged under the modular flow \cite{longo1982algebraic}. Enlarging the algebra of QFT by the modular group factors the local algebra of QFT completely: $\mH=\oplus_q \mK_{q}^A\otimes \mK_{q}^{A'}$ where $q$ is the modular frequency. The enlarged algebra is type II$_\infty$ and has a trace \cite{takesaki1973duality}. Entanglement entropy in the extended Hilbert space is divergent, however the factorization of the Hilbert space resembles the structure of boundary quantum field theory, and the insertion of a resolution of identity that is the center $\oplus_q \ket{q}\bra{q}$ in the algebra.\footnote{We thanks James Sully for pointing out this connection to us.}

\section{Conclusions}

In this work, we generalized the definition of entanglement entropy to the cases with no tensor product structure, and used the new definition to define an entanglement measure that captures the contribution of charges to entanglement in quantum systems with symmetries in equation (\ref{final2}). 
The proposed measure is comprised of two relative entropies. One is the relative entropy with respect to the charge neutral operators  and the other is the relative entropy due to the charge creation operators. We used representation theory to introduce the charge creation operators called intertwiners and bi-local intertwiners, and wrote down relative entropy that capture their contributions to entanglement. We set up the formalism to compute these measures in QFT using the Tomita-Takesaki modular theory. We highlighted the differences between QFT and lattice models, and discuss an extension of the algebra of QFT that leads to a factorization of the charged modes.

We would like to thank Horacio Casini,  Roberto Longo, Thomas Sinclair, James Sully and Edward Witten for many valuable discussions. NL would also like to thank the Institute for Advanced Study for their hospitality during his visit. Funding for this visit was provided by the NSF grant PHY-1911298.

\appendix

\section{Group and algebra extensions}\label{app:A}

\subsection{Group extension: semi-direct product}

Given two groups $N$ and $H$ consider the trivial extension that is the Cartesian product group $N\times H$ where elements of group are $(n,h)$ and the multiplication is $(n_1,h_1).(n_2,h_2)=(n_1n_2,h_1h_2)$. If $H$ acts on $N$ by an outer automorphism $\phi_h: n\to h n h^{-1}$ with the composition rule $(\phi_{h_1}\circ\phi_{h_2})(n)=\phi_{h_1h_2}(n)$   we can consider a subgroup of $G=N\rtimes_\phi H\subset N\times H$ called the semi-direct product and has the multiplication rule $(n_1, h_1).(n_2,h_2)=(n_1 \phi_{h_1}(n_2), h_1 h_2)$. The inverse of $(n,h)$ is $(\phi_{h^{-1}}(n^{-1}),h^{-1})$. All we need for the construction of the semi-direct product is the homomorphism $\phi:H\to \text{Aut}(N)$.

In the semi-direct product extension $G$, $N$ is a normal subgroup and $H=G/N$ is the quotient group.
An important example is the Poincare group that is the semi-direct product of translations and the Lorentz group: $\mathbb{R}^{1,d-1}\rtimes O(1,d-1)$. If $N$ is the center of $G$ the semi-direct product is called a central extension. A trivial example of central extension is the direct product group $N\times H$ where $N$ is Abelian. Non-trivial examples comes from the study of the projective representations of a group. Consider a group $H$, the Abelian group of complex numbers $\mathbb{C}$ and the map $\phi_h(\alpha)=\alpha c(\alpha,h)$ with $c(\alpha,h)$ a complex number. If $c(\alpha,h)c(\beta,h)=c(\alpha\beta,h)$ and $c(\alpha^*,h)=c(\alpha,h)^*$, this map is an outer automorphism of $\mathbb{C}$, and we can construct $\mathbb{C}\rtimes_
\phi H$ with the multiplication rule $(\alpha,h_1).(\beta,h_2)=(\alpha\beta c(\beta,h_1),h_1h_2)$. We need to further check that $\phi_{h_1}(\alpha)\phi_{h_2}(\alpha)=\phi_{h_1h_2}(\alpha)$ which imposes $c(\alpha,h_1)c(\alpha,h_2)=c(\alpha,h_1h_2)$.

\subsection{Lie algebra extension: semi-direct sum}

Consider the groups $H$ and $N$ are Lie groups and their corresponding Lie algebras $\mathfrak{h}$ and $\mathfrak{n}$. The map $\phi:H\to \text{Aut}(N)$ induces a map $\psi:\mathfrak{h}\to \text{Aut}(\mathfrak{n})$ defined by the Lie correspondence
\begin{eqnarray}
\psi_{\hat{h}}(\hat{n})=\frac{d}{dt}\lb \phi_{e^{t \hat{h}}}(e^{t\hat{n}})\rb_{t=0}
\end{eqnarray}
where $\hat{n}$ and $\hat{h}$ are elements of the Lie algebra $\mathfrak{n}$ and $\mathfrak{h}$, respectively. 
We obtain the notion of a semi-direct sum of Lie algebras with the Lie bracket defined using the equation
\begin{eqnarray}
\left[(\hat{n}_1,\hat{h}_1),(\hat{n}_2,\hat{h}_2)\right]=([\hat{n}_1,\hn_2]+\psi_{\hh_1}(\hn_2)-\psi_{\hh_2}(\hn_1),[\hh_1,\hh_2])\ .
\end{eqnarray}

There is another method to centrally extend Lie algebras. Every linear map $\chi:\mathfrak{h}\times \mathfrak{h}\to \mathbb{C}$ that is anti-symmetric, i.e. $\chi(\hat{h}_1,\hat{h}_2)=-\chi(\hat{h}_2,\hat{h}_1)$, and satisfies the Jacobi identity leads to an extension defined by the Lie bracket
\begin{eqnarray}
\left[(\alpha, \hat{h}_1), (\beta,\hat{h}_2) \right]=( \chi(\hat{h}_1,\hat{h}_2), [\hat{h}_1,\hat{h}_2])\ .
\end{eqnarray}

A finite-dimensional simple Lie algebra has no non-trivial central extensions. To find examples of non-trivial central extension we have to consider infinite-dimensional Lie algebras. As an example, we work out the central extension of the polynomial loop algebra: Kac-Moody algebra.
The loop group is defined to be the algebra of smooth $G$-valued functions on a circle with group multiplication rule. These are loops $C$ on the group $G$, $C: S^1\to G$ with $(C_1C_2)(\theta)=(C_1)(\theta)(C_2)(\theta)$. A loop Lie algebra is the vector space of smooth functions from $S^1$ to $\mathfrak{g}$ of $G$.

Consider the tensor product space $\mathfrak{g}\otimes C^\infty(S^1)$, where $\mathfrak{g}$ is a finite dimensional simple Lie algebra and $C^\infty(S^1)$ is the algebra of smooth functions on $S^1$. This vector space is a Lie algebra with the bracket defined by 
\begin{eqnarray}
[\hat{g}_1\otimes f_1,\hat{g}_2\otimes f_2]=[\hat{g}_1,\hat{g}_2]\otimes f_1f_2\ ,  \quad (\hat{g}_1,\hat{g}_2 \in \mathfrak{g}) .
\end{eqnarray}
Importantly, this space is not a direct product of the two spaces $\mathfrak{g}$ and $C^\infty(S^1)$ due to the smoothness condition of functions. Instead, it should be thought of as the Lie algebra of smooth $\mathfrak{g}$-valued functions of $S^1$. The Fourier transform on $S^1$ gives the basis $\hat{g}\otimes e^{in \theta}$ where $\theta$ is the angle on $S^1$ and $n\in\mathbb{Z}$. The Lie algebra generated by such generators is the polynomial loop algebra. 
Another way to think about this algebra is in terms of  the algebra of Laurent polynomials $\sum_{n\in\mathbb{Z}}f_n z^n$ with only finitely many non-zero $f_n$ and the standard multiplication and addition. Then, the algebra of $G$-valued functions on $S^1$ is the Lie algebra of formal sums $\sum_{n\in \mathbb{Z}} z^n \otimes \hat{g}_n$ with the Lie bracket
\begin{eqnarray}
[z^n\otimes \hat{g}_1,z^m\otimes \hat{g}_2]=z^{n+m}\otimes [\hat{g}_1,\hat{g}_2]\ .
\end{eqnarray}
The generators of the Lie algebra $J_a$ satisfy
\begin{eqnarray}
[z^n\otimes \hat{J}_a,z^m\otimes \hat{J}_b]=\sum_c C^c_{ab} z^{n+m}\otimes \hat{J}_c\ .
\end{eqnarray}
where $C^c_{ab}$ denotes the structure constants of the Lie algebra $\mathfrak{g}$. The central extension of this algebra is $\mathfrak{g}\otimes C^\infty(S^1)\oplus \mathbb{C}$
\begin{eqnarray}
[(\alpha, z^n\otimes \hat{J}_a),(\beta,z^m\otimes \hat{J}_b)]=(k\, n\, K(\hat{J}_a,\hat{J}_b)\delta_{n+m,0},\sum_k C^c_{ab} z^{n+m}\otimes \hat{J}_c)\ .
\end{eqnarray}
where $K(J_a,J_b)$ is the Killing form on $\mathfrak{g}$ and $k$ is the central charge. This is an affine Lie algebra.

\subsection{von Neumann algebra extension: Crossed product}

Groups can act on von Neumann algebras and one can extend an algebra $\mA$ by a group $G$ that acts on it as outer automorphisms to obtain a larger algebra called the crossed product $\mA\rtimes_\phi G$ \cite{jones2015neumann,takesaki2013theory}. 
If the action of the $G$ on $\mA$ is $\phi_g(a)=a_g=u_g a u_g^{-1}$ with $u_gu_h=u_{gh}$ we add $u_g$ to the set of operators in our algebra and consider the algebra of formal sums $\sum_{g\in G}a_gu_g$ with $a_g\in\mA$. 
If $\mA$ acts on the Hilbert space $\mH$ and $L^2(G)$ is the Hilbert space of square-integrable functions of the group the crossed product algebra acts on $\mH\otimes L^2(G)$; that is the space of square-integrable $\mH$-valued functions of $G$. Vectors of this Hilbert space are $\ket{\Psi}=\sum_{g\in G}c_g\ket{\Psi;g}$ and the inner product is 
\begin{eqnarray}
\braket{\Psi|\Phi}=\sum_{g\in G} c_g^*b_g \braket{\Psi;g|\Phi;g}\ .
\end{eqnarray}
The multiplication rules are $u_h\ket{\Psi;g}=\ket{\Psi;h g}$ and $a_g\ket{\Psi;g}=\ket{a\Psi;g}$.

\subsection{Dual group and non-Abelian Fourier transform}

Consider a locally compact Abelian group $G$. The characters of $G$ are linear maps from $G$ to complex numbers. For instance, for the group $U(1)$ of rotations on a circle the characters are $\chi(U_\theta)=e^{i \theta \chi}$ with $\theta\in [0,2\pi)$. The point-wise multiplication $(\chi_1 \chi_2)(U_\theta)=\chi_1(U_\theta)\chi_2(U_\theta)$ gives the characters the structure of a group called the {\it dual group} of $G$ that we denote by $\hat{G}$. The dual group allows us to define a Fourier transform for functions on the group $G$: 
\begin{eqnarray}
\hat{f}(\chi)=\sum_{g\in G} \chi^{-1}(g) f(g)\ .
\end{eqnarray}
If the group is finite the dual Fourier transform is 
\begin{eqnarray}
f(g)=\frac{1}{|G|}\sum_{\chi \in \hat{G}}\chi(g) \hat{f}(\chi)\ .
\end{eqnarray}
To generalize Fourier transform to non-Abelian finite groups $G$ we replace the character of group with its irreducible representations $\rho_r(g)$:
\begin{eqnarray}
\hat{f}(\rho_r)=\sum_{g\in G} \rho_r(g) f(g)\ .
\end{eqnarray}
If $\rho_r(g)$ is represented by a $d_r\times d_r$ matrix then $\hat{f}(g)$ is also a matrix of same dimensions. The inverse Fourier transform is 
\begin{eqnarray}
f(g)=\frac{1}{|G|}\sum_r d_r\text{tr}\lb\hat{f}(\rho_r)\rho_r(g^{-1})\rb
\end{eqnarray}
where the sum is over irreducible representations $\rho_r$ of group $G$ and we have used the fact that $\frac{1}{|G|}\sum_r d_r \text{tr}\lb\rho_r(g)\rb = \delta_{g\mathbb{I}}$ \cite{etingofRepTheory}. The analog of the multiplication of characters in the non-commutative case is the tensor product of irreducible representations which does not form a group, because the tensor product of irreducible representations is not in general irreducible.



\bibliographystyle{JHEP}
\bibliography{main}

\end{document}